\documentclass[acmsmall,table,xcdraw]{acmart}
\AtBeginDocument{%
  }

\newenvironment{interviewquote}{%
    \begin{quote}%
    \itshape
    }{%
    \end{quote}%
}

\usepackage{url}
\usepackage{caption} 
\setcopyright{cc}
\setcctype{by}
\acmJournal{PACMHCI}
\acmYear{2025} \acmVolume{9} \acmNumber{7} \acmArticle{CSCW264} \acmMonth{11} \acmDOI{10.1145/3757445}

\received{July 2024}
\received[revised]{December 2024}
\received[accepted]{March 2025}

\begin{document}

\title{``There Has To Be a Lot That We're Missing'': Moderating AI-Generated Content on Reddit}

\author{Travis Lloyd}
\email{tgl33@cornell.edu}
\orcid{0009-0009-7393-7105}
\affiliation{%
  \institution{Cornell University}
  \department{Cornell Tech}
  \city{New York}
  \state{NY}
  \country{USA}
}

\author{Joseph Reagle}
\email{j.reagle@northeastern.edu}
\orcid{0000-0003-0650-9097}
\affiliation{%
  \institution{Northeastern University}
  \department{Communication Studies}
  \city{Boston}
  \state{MA}
  \country{USA}
}

\author{Mor Naaman}
\email{mor.naaman@cornell.edu}
\orcid{0000-0002-6436-3877}
\affiliation{%
  \institution{Cornell Tech}
  \city{New York}
  \state{NY}
  \country{USA}
}

\begin{abstract}
Generative AI is altering how we work, learn, communicate, and participate in online communities. 
How might online communities be changed by generative AI? 
To start addressing this question, we focused on online community moderators' experiences with AI-generated content (AIGC). 
We performed fifteen in-depth, semi-structured interviews with moderators of Reddit communities that restrict the use of AIGC.
Our study finds that rules about AIGC are motivated by concerns about content quality, social dynamics, and governance challenges.
Moderators fear that, without such rules, AIGC threatens to reduce their communities' utility and social value.
We find that, despite the absence of robust tools for detecting AIGC, moderators were able to somewhat limit the disruption it caused by working with their communities to clarify norms. 
However, moderators found enforcing AIGC restrictions challenging, as they rely on time-intensive and inaccurate detection heuristics.
Our results highlight the importance of supporting community autonomy and self-determination in the face of this sudden technological change, and suggest potential design solutions that may help. 
\end{abstract}

\ccsdesc[500]{Human-centered computing~Empirical studies in collaborative and social computing}

\keywords{Online Communities, Reddit, AI-Generated Content, Generative AI, Moderation}

\maketitle

\section{Introduction}
Online communities are adjusting their policies and practices to address the proliferation of AI-generated content~(AIGC)~\cite{Thompson23Discovered} produced by generative AI~\cite{OpenAI22,ramesh2021zeroshot}. 
The presence of AIGC, whether disclosed or not, may alter the social dynamics of online communities. 
For example, AI-generated text with subtle inaccuracies may strain collaborative knowledge-sharing communities, such as Stack Exchange~\cite{makyen23} and Wikipedia~\cite{gertner_wikipedias_2023}. 
Other communities, like Facebook groups where people gather for human connection~\cite{fiesler_ai_2024}, might see AIGC as inauthentic and disruptive.
This new form of disruption adds to online communities' existing challenges, such as harassment or coordinated disinformation campaigns, which generative AI threatens to make cheaper, more effective, and more personalized~\cite{goldstein23,woodDarkEchoesExploitative2024}.

As many communities rely on content moderators for stewardship~\cite{grimmelmann2015virtues}, it is important to understand and document moderators' firsthand experiences responding to AIGC.
Effective moderation can help communities clarify and enforce community norms~\cite{Kiene16,kiesler11,SeeringEtal2019mec} in order to stay civil in times of disruptive change. 
Existing moderation strategies will likely need to adapt in order to remain effective in the face of AIGC.
Indeed, there is evidence from recent quantitative research that online communities are enacting rules about the use of AIGC~\cite{lloydAIRulesCharacterizing2025}, but exactly \textit{how} and \textit{why} they implement such rules requires further study.

In order to more deeply understand generative AI's potential impact on online communities, this paper qualitatively investigates the experience of moderating AIGC on Reddit.
Not all online communities see AIGC as a concern, but we center those that do to illuminate how platform designers and community stewards might help such communities navigate moments of dramatic technological change~\cite{Seering20}.
How to provide such support is a natural question for the HCI and Social Computing research communities, 
a central goal of which has been to encourage healthy communities as they strive to meet a variety of individual and collective needs~\cite{kiesler11,kraut11,Lampe10}. 

To this end, we conducted fifteen semi-structured interviews with volunteer Reddit moderators. 
We focus on Reddit, a social sharing and news aggregation site, to compare and contrast several relatively independent communities that, nonetheless, share a platform. 
Reddit is a popular platform to study~\cite{proferes21} because of its variety of communities and distributed governance structure: subreddits set and enforce their own rules to supplement site-wide policy~\cite{FieslerEtal2018rrc}.
Studying Reddit thus allows us to explore how different attitudes and practices towards AIGC organically emerge across communities.
We qualitatively explore moderators' perspectives on what constitutes AIGC and what makes it problematic.
Rather than try to provide a representative view of all Reddit moderators or communities, our goal is to better understand the concerns of moderators in communities that have restricted the use of AIGC.
We conducted our interviews within a year of the launch of ChatGPT.
While online communities' concerns about AIGC have continued to evolve~\cite{lloydAIRulesCharacterizing2025}, our findings highlight the early concerns of moderators forced to adapt to a new technology.
Our work thus explores the following research questions:
\begin{itemize}
    \item \textbf{RQ1:} What are volunteer moderators' attitudes towards AIGC in online communities that restrict its use?
    \item \textbf{RQ2:} What experiences have these moderators had with AIGC in their communities?
\end{itemize}

Our exploration of \textbf{RQ1} reveals that our participants restrict the use of AIGC because they are concerned that it may negatively impact the content quality, social dynamics, and governance processes in their communities.
Though our participants acknowledge potential benefits of AIGC, they generally oppose its use---some even see AIGC as completely antithetical to their communities' purposes and values.
Our participants shared several specific threats from AIGC.
Participants say that AIGC is poorly written, inaccurate, and off-topic, and thus threatens to reduce the quality of their communities' content.
Participants also report that AIGC may harm community social dynamics by reducing opportunities for human connection, straining relationships, and violating shared values.
Finally, participants are concerned that AIGC may increase the scale and sophistication of existing types of problematic behavior, making the jobs of moderators more difficult.

Our exploration of \textbf{RQ2} reveals that while some participants feel that by clarifying rules and norms they have made AIGC no more threatening than other problematic content, AIGC has increased the workload on already overburdened moderators. 
Rules about AIGC are hard to enforce: moderators believe they can detect AIGC, but the process relies on time-consuming and imperfect heuristics that look for certain ``tells'' in content and behavior.
Moderators sometimes use automated detection tools to help with enforcement, but none are reliable enough to fully automate the process. 
Additionally, moderators acknowledge the sensitivity of accusing community members of posting AIGC and emphasize the care that goes into such decisions.

We discuss the implications of these findings for members of online communities, community moderators, and platform designers.
We explore the relationship between community values and attitudes towards AIGC and suggest that the arrival of AIGC will drive communities to develop new ways to demonstrate authenticity. 
We discuss platform designs that can support communities by encouraging authentic communication, as well as others that can keep members aware of evolving community norms.
We also discuss potential bias in moderators' detection methods and explore design solutions that may improve them by synthesizing relevant information sources.
Finally, to preserve the types of interactions that communities value, we suggest that platforms allow communities to opt out of any generative AI features.

\section{Related Work}
Understanding the challenges posed by AIGC requires an understanding of online communities, their health, and the challenges of moderation.
Below we summarize literature showing that online communities vary in their values and their reasons for forming, suggesting that community attitudes towards AIGC will be context-dependent.
The literature on community health highlights the challenges of dealing with deception, low engagement, and harassment, all of which might be amplified by AIGC.
Finally, we summarize research that emphasizes the challenges of content moderation; the difficulty of identifying and addressing AIGC might further strain moderators and their relationships with their communities.

\subsection{Online Community Types and Values}
Online communities form for different reasons, have different values, and thus may feel threatened by AIGC in different ways.
Early work by Preece~\cite{preece_empathic_1999} coined the term \emph{empathic communities} to refer to sites of both emotional and factual communication, where member participation is motivated by a desire for something that professionals cannot provide: authentic, firsthand accounts from others with similar lived experiences.
Recent work has shown that these communities consider certain types of interactions inauthentic that are completely acceptable in other contexts~\cite{Smith23}.
We suspect that empathic communities may see AI-written posts as problematic due to their lack of lived experience.
Another useful concept comes from the work of Lave and Wegner, who use the term \emph{community of practice}~\cite{lave1991situated} to denote a group of people united by a desire to improve their skills by doing, sharing, and receiving feedback from one another.
These communities care deeply about process and might regard contributions made with generative AI as inauthentic.
Finally, Jenkins' concept of \emph{knowledge communities}, in which ``members work together to forge new knowledge, often in realms where no traditional expertise exists''\cite{jenkins_introduction_2006}, is very relevant to our study.
Knowledge communities value accurate information and expertise, which may be threatened by generative AI's current tendency to produce inaccurate ``hallucinations''~\cite{augenstein2023factuality}. 
Kraut and Resnick's work supplements this conceptual literature with practical considerations for platform designers~\cite{kraut11}, including ways to build systems that support reputation and trust~\cite{resnick2002trust}.
These systems may need to evolve in a world where contributions made with AI are indistinguishable from those produced by experts.

Reddit has been a frequent topic of study for empirical investigations into the different types and values of online communities~\cite{proferes21}.
Reddit's hierarchical governance structure, in which site-wide policies are supplemented by community-derived and enforced rules, allows it to host many independent communities with distinct purposes and values.
This model of community self-moderation~\cite{Seering20} allows researchers to study emergent attitudes and practices as they evolve.
For example, recent work has shown how the use of governance bots shapes Redditors' experience of a community, primarily by affecting their sense of virtual community~\cite{Smith22}. 
Other research has used surveys of community members~\cite{weld_what_2022}, analysis of public subreddit rules~\cite{FieslerEtal2018rrc,lloydAIRulesCharacterizing2025}, and logs of moderation actions~\cite{Chandrasekharan18} to study community values across the platform.
These studies emphasize the diverse set of values and rules held by different subreddits. 
Though some are common, such as rules to discourage harassment, many are context-dependent. 
For example, Fiesler et al~\cite{FieslerEtal2018rrc} found that subreddit topic is more predictive of subreddits' rules than size.
Recent work introduced the framework of \textit{Community Archetypes} to describe common types of communities on Reddit~\cite{Prinster24}, which we use to make sense of the variation in community stances toward AIGC. 
Using this framework, our recent quantitative analysis~\cite{lloydAIRulesCharacterizing2025} shows that \textit{Content Generation} communities are the most likely to have rules about AI, while \textit{Social Support} communities are negatively associated with such rules.
On the whole, the findings from this literature suggest that subreddits will not have a unified stance towards AIGC; consequently, we seek to understand different moderators' perspectives.

\subsection{Generative AI and Threats to Healthy Online Communities}
The HCI literature has identified varied threats to online communities that generative AI might amplify, such as deception, low engagement, and harassment.
Past studies have shown online communities to be vulnerable to strategic influence operations, and recent speculative work has hypothesized that generative AI will increase the volume and believability of such efforts~\cite{augenstein2023factuality,goldstein23}.
Indeed, early empirical studies have found AIGC to be as or more deceptive than human-written content~\cite{park_ai_2024}, and identified AIGC in fake online reviews~\cite{Oak24} and scams~\cite{diresta2024spammers}.
Other studies have focused on the deception capabilities of algorithmically controlled social media accounts, or \textit{social bots}.~\cite{Ferrara2016,menczer_addressing_2023}.
While LLM-powered social bots have already been detected on Twitter/X~\cite{yang23}, our work looks beyond social bots to include situations where \emph{humans} use generative AI tools to create content.
We hypothesize that humans posting undisclosed AIGC may be regarded as a more subtle form of deception.
Findings from AI-Mediated Communication (AIMC) studies suggest that AI affects interpersonal communication~\cite{hancock_aimc_2020}, causing people to regard others more negatively~\cite{rae_effects_2024, kadoma2024generativeaiperceptualharms} and rate them as less trustworthy~\cite{Jakesch19} when they perceive that the other is using AI. 

Another threat to online communities is that of declining participation.
The HCI and CSCW literature provides techniques to encourage community contribution~\cite{kraut11} and commitment~\cite{ren_encouraging_2012}, such as by establishing a group identity that can encourage strong ties between members.
However, specifically studying how \textit{AIGC} affects participation in online communities is a relatively new area of research~\cite{wei2024understandingimpactaigenerated}.
Several studies of sites in the Stack Exchange network have demonstrated that the launch of ChatGPT caused an overall decline in website visits, unique users, and question volume, while increasing the complexity of questions asked~\cite{burtch_consequences_2024,sanatizadeh_exploring_2023,xue_can_2023}.
The authors attribute these effects to the relative ease and speed with which users can get simple questions answered by LLM-powered chatbots off-platform. 
Interestingly, the authors of one study~\cite{burtch_consequences_2024} contrast this effect with a measured null-effect on similar Reddit communities.
The authors suggest that these Reddit communities avoid a decline in activity due to their emphasis on socialization, rather than pure information exchange. 
Recent work has also studied the effects of top-down bans on AIGC and found that banning the use of generative AI on Stack Exchange~\cite{wang_can_2024} led to a decrease in volume and quality of questions and answers.
Our study builds on this line of inquiry by asking if the emergence of AIGC and consequent bans threaten participation in self-governing communities with varied values and needs.

Finally, online communities are often sites of harassment and abuse, a frequent topic in the CSCW literature~\cite{Cheng17,han_hate_2023};
the effects of AI-powered variants, such as malicious synthetic, or deepfake, media~\cite{farid_creating_2022,rini_deepfakes_2022} are newer and much less studied.  
Given the potential for AI-powered abuse, our work explores how community moderators are thinking about these threats and what they are doing to counter them.

\subsection{The Challenges of Volunteer Community Moderators}
Subreddits that decide to enact rules about AIGC will turn to already-strained volunteer moderators to enforce these rules~\cite{li_all_2022,Wohn19}. 
Moderation involves both ``visible'' and ``invisible'' work~\cite{Gilbert20}, and is alternatively characterized as ``civic labor''~\cite{Matias19}, ``emotional labor''~\cite{Dosono19}, and ``care work''~\cite{gilbert23}.
The HCI community has a long history of addressing the needs of moderators~\cite{Jhaver19}.
Though existing research has focused on handling harmful content such as misinformation~\cite{Bozarth2023}, the specific challenges of moderating AIGC have not been studied.

One perpetual challenge for moderators is maintaining alignment with their community members~\cite{Koshy23}. 
Given the power dynamic between moderators and regular members, perceptions of unfair or biased moderation decisions can strain community relationships.
Perceptions of bias are not unfounded: existing work has shown that individuals with marginalized identities are more likely to be the target of moderation actions~\cite{Haimson21,Thach24}.
We suspect this will be a problem when moderators enforce rules about AIGC, as research has found that humans cannot reliably distinguish between AI and human-generated text~\cite{clark21} or images~\cite{nightingale22}, and use flawed heuristics to make such determinations~\cite{Jakesch_2023}.
Though little is known about moderators' AI detection practices, recent work has shown that in general, determinations about AI use are biased according to race and gender stereotypes~\cite{mink24}.
Our study builds on this work by exploring the real-world AI detection practices of community moderators.

The CSCW research community has researched and developed computational tools meant to aid moderators in their work~\cite{Chandrasekharan19,Schluger22,Zhang20}.
However, detecting AIGC with computational techniques has, so far, proven challenging~\cite{yang_survey_2023}. 
While commercial classifiers for AI-generated text claim to be effective~\cite{Crothers23}, their performance has been questioned by the research community~\cite{sadasivan2023}.
Recent work has found these classifiers to perform no better than random chance~\cite{weberwulff23}, or even worse, to be biased against the writing of non-native English speakers~\cite{liang23}.
This calls into question the feasibility of a reliable, general-purpose detection tool that will work on any snippet of text. 
Still, other studies have had success identifying AIGC in specific types of writing, such as news articles~\cite{hanley_machinemade_2024}, conference peer reviews~\cite{liang_monitoring_2024}, crowd-worker output~\cite{veselovsky_prevalence_2023}, and scientific paper abstracts~\cite{kobakDelvingLLMassistedWriting2025}, suggesting that such techniques might be adapted to other contexts.
Our study will investigate how moderators think about these tools and currently use them in their work.

\section{Methods: Interview Study}
We chose to conduct semi-structured interviews with Reddit moderators, approaching our research questions qualitatively, to deeply ``explore and understand the meaning individuals or groups ascribe to a social or human problem''~\cite{creswellResearchDesignQualitative2017a}, specifically the newly introduced challenges of AIGC.
We identified target communities by looking at their public stances towards AIGC, and recruited participants via outreach and snowball methods.

\subsection{Recruitment}\label{recruitment}
As our research goal was to explore moderator concerns, we sought to speak with moderators of communities who had restricted the use of AIGC.
To identify such communities, we turned to public data about subreddits' community rules.
We performed a crawl of the Reddit website in the summer of 2023 and extracted the text-based rules from all subreddits indexed by Reddit's ``Top Communities'' list\footnote{\url{https://www.reddit.com/best/communities/1/}}, which includes public subreddits with more than ten subscribers.
We crawled $n=337,399$ subreddits and found $n=87,596$ with non-empty rules and English as a primary language.
We then performed a text search on the rules to identify subreddits with rules related to AIGC. 
Though this method may be imprecise, it was sufficient for our purpose, which was to understand broad patterns to guide our recruitment strategy.

In the data that we gathered, 4\% of all subreddits had rules about AI, but these rules were much more common in larger subreddits: when we restricted our analysis to the top 1\% of subreddits by subscriber count, 20\% had rules governing the use of AI.
We thus designed a recruitment strategy focused on the largest subreddits, which appear to be the most concerned about AIGC.
Many moderators of larger subreddits also moderate small forums and were able to speak about differences of scale.

We recruited and interviewed participants during the summer of 2023.
We manually examined the rules of subreddits in descending order of subscriber count.
When we found a subreddit with either an explicit rule governing the use of AI or a rule that might implicitly cover AI use, we added it to our sample.
We also looked for moderator discussions about AIGC in other public forums, such as the r/ModSupport subreddit, and added those moderators' subreddits to our sample.
We then messaged the moderators of each subreddit in our sample via Reddit's Modmail\footnote{\url{https://support.reddithelp.com/hc/en-us/articles/210896606-What-s-mod-mail-and-how-do-I-send-a-message}} 
 feature, 
sending Modmail recruitment messages to 60~subreddits.
We used two additional techniques to recruit participants beyond the largest subreddits.
First, we used snowball sampling and asked moderators to forward our recruitment message to other potential participants.
Second, we posted our recruitment call on several social media platforms.
To ensure respondents were actual moderators, we asked for their Reddit username and the subreddit that they moderate, then verified that they were indeed listed as moderators on their Reddit user profile.
Participants were offered \$20 gift cards as compensation.
We conducted interviews on a rolling basis and continued recruiting until we had enough rich and complex data to adequately address our research questions, as per Braun and Clarke's saturation criteria for Reflexive Thematic Analysis~\cite{braunSaturateNotSaturate2021}.
Overall, we interviewed fifteen participants, meeting or even exceeding the sample size norms of the HCI research community~\cite{caine16}.
These participants collectively moderated over~100 different subreddits with sizes ranging from less than ten members to more than 32~million (see participant information in Table~\ref{tab:participants}). 

\subsection{Interviews}
We designed an initial interview protocol based on our research questions and updated it as themes emerged during early interviews.
Since pseudonymity is an important feature of the Reddit platform, we followed best practices in ethical online communities research~\cite{Fiesler24} and took special considerations to protect the privacy of our participants.
We did not ask for or record participants' names, demographic information, or any offline identifiers.
As is common in Reddit and Social Computing research~\cite{Gilbert20,Jhaver19}, we gave participants the opportunity to interview via text-based mediums in order to increase participation and representation.
Text is the primary mode of interaction on Reddit, so despite the fact that such interviews may produce shorter exchanges and less rich insights than oral interviews~\cite{Bruckman2006-BRUTST,Dimond12}, we believed they would be a valuable supplement to our oral interviews.
Accordingly, we allowed participants to choose to interview via synchronous video or audio call, semi-synchronous message exchange on the Reddit platform, or asynchronous email exchange.
Regardless of channel, we posed our research questions one at a time and followed up in a semi-structured way.
For example, for text-based interviews, we were online at the same time as our participants, exchanging messages in a back-and-forth manner, which allowed us to probe participants' responses as necessary for deeper insight or clarity.

Our study was approved by the Cornell University IRB. 
All participants read and accepted a consent agreement permitting their responses and subreddit affiliations to be published in research reports.
To ensure that we reliably represented participants' views, we contacted participants for feedback and approval on their included quotes before publication, following CSCW community recommendations~\cite{mcdonald19}. 
We integrated all participants' feedback in the final manuscript.

The first author conducted the interviews using these key guiding questions:
\begin{enumerate}
    \item What types of AIGC are moderators seeing in their communities?
    \item What do moderators think are the motives of AIGC posters? 
    \item How are moderator and non-moderator community members responding to AIGC?
    \item What concerns do moderators have about AIGC and how do they compare to other moderation concerns?
    \item How do moderators identify AIGC?
    \item What could help moderators address these concerns?
\end{enumerate}

Transcripts for interviews conducted via audio and video calls were generated automatically from recordings, then manually reviewed and corrected by the first author. 
For interviews done via email and Reddit message, the text was captured in similarly formatted transcript files for analysis. 
Interviews lasted between 26 minutes and 147 minutes with a mean and median of 71 and 62 minutes, respectively.
While we scheduled $60$ minutes for interviews, some oral interviews ended early and some text-based interviews went longer, as participants took their time to respond to our questions: oral interviews had a mean duration of $47$ minutes ($SD=15$), compared to a mean duration of $91$ minutes ($SD=38$) for written interviews.
Despite the non-synchronous nature of the text-based interviews, all of these interviews yielded meaningful conversations with at least eight back-and-forth question and answer exchanges.  
Text-based interviews provided rich data that was not shared in oral interviews, such as links to example AIGC posts from participants' communities.
Each interview medium yielded insights that, when combined, yield a more nuanced picture than any single medium could provide.

The first two authors then collaborated to iteratively code the interview content into categories, using an inductive thematic analysis method inspired by Braun and Clarke's Reflexive Thematic Analysis~\cite{braun21,Braun19}.
The two authors began by discussing themes that the first author observed during interviews and transcript correction.
From this conversation, an initial high-level set of codes emerged.
Next, the first author open-coded all of the interview transcripts at the sentence-level, applying the initial codes and creating new ones where existing codes were insufficient.
The two authors then reviewed the set of codes and refined them by merging similar codes and nesting those within a hierarchy.
This round of coding did not produce new themes, only new sub-codes, suggesting that the themes were fairly stable. 
Both authors then independently applied these codes to the same three interview transcripts and compared their results.
Neither created any new codes and the few cases of disagreement were resolved by further merging similar codes. 
This process yielded approximately 400 tagged phrases across all transcripts, which were grouped into the three high-level themes that we discuss below.

\begin{table*}
\scriptsize
\caption{Details for each interview. For each participant, we note the largest subreddit (according to subscriber count at the time of the interview) and the number of individual subreddits that they moderated.}
\begin{tabular}{llllll}
\arrayrulecolor{gray}\hline
\textbf{Interviewee} & \textbf{\begin{tabular}[l]{@{}l@{}}Largest Moderated Subreddit\\  (Subscribers)\end{tabular}} & \textbf{\begin{tabular}[l]{@{}l@{}}Subreddits \\ Moderated\end{tabular}} & \textbf{\begin{tabular}[l]{@{}l@{}}Interview \\ Medium\end{tabular}} & \textbf{Primary Threat}                         & \textbf{\begin{tabular}[l]{@{}l@{}}Concern \\ Level\end{tabular}} \\ \hline
1                    & r/todayilearned (32M)                                                                              & 25                                                                       & Audio Call                                                           & Amplified Attacks                        & High                                                              \\ 
2                    & r/food (23M)                                                                                       & 25                                                                       & Reddit Messages                                                      & Lowered Quality                & High                                                              \\ 
3                    & r/explainlikeimfive (23M)                                                                          & 3                                                                        & Email Exchange                                                                & Less Accurate Information                 & High                                                              \\ 
4                    & r/explainlikeimfive (22M)                                                                          & 2                                                                        & Audio Call                                                           & Less Accurate Information                 & High                                                              \\ 
5                    & r/WritingPrompts (17M)                                                                             & 3                                                                        & Video Call                                                           & Less Skill Development          & High                                                              \\ 
6                    & r/politics (8.3M)                                                                                  & 25                                                                       & Reddit Messages                                                      & Strained Relationships                           & Medium                                                            \\ 
7                    & r/CryptoCurrency (6.7M)                                                                            & 8                                                                        & Reddit Messages                                                      & Amplified Attacks                        & Medium                                                            \\ 
8                    & r/itookapicture (5.1M)                                                                             & 2                                                                        & Email Exchange                                                               & Less Skill Development           & Medium                                                            \\ 
9                    & r/NintendoSwitch (5.0M)                                                                            & 23                                                                       & Email Exchange                                                               & Lowered Quality                & Medium                                                            \\ 
10                   & r/Fantasy (3.4M)                                                                                   & 1                                                                        & Email Exchange                                                               & Lowered Quality                & Medium                                                            \\ 
11                   & r/changemyview (3.4M)                                                                              & 2                                                                        & Video Call                                                           & Less Human Connection                & Low                                                               \\ 
12                   & r/changemyview (3.4M)                                                                              & 2                                                                        & Audio Call                                                           & Less Human Connection & Low                                                               \\ 
13                   & r/AskHistorians (1.8M)                                                                             & 4                                                                        & Video Call                                                           & Less Accurate Information                 & Medium                                                            \\ 
14                   & r/AskHistorians (1.8M)                                                                             & 1                                                                        & Reddit Messages                                                      & Less Accurate Information                 & Medium                                                            \\ 
15                   & r/GCTrading (22.7K)                                                                                & 3                                                                        & Audio Call                                                           & Amplified Attacks                        & Low                                                               \\ \hline
\end{tabular}
\label{tab:participants}
\end{table*}

\section{Findings}\label{definitions}
Three main themes emerged from our interviews.
With regards to RQ1, our interviews surfaced insights into moderators' (1) concerns about AIGC in their communities; answering RQ2, our interviews elicited details on (2) how moderators' communities are responding to AIGC and (3) the challenges of enforcing AIGC rules.

Our interviews did not reveal a single definition of ``AIGC''. Instead, our participants mentioned three different categories of content related to our inquiry: images produced by generative AI, text produced (entirely or partially) by generative AI, and posts made by automated accounts (bots). 
Communities sometimes banned one of these categories while allowing others.

Participants generally used the term ``AI-generated'' in a pejorative way, but some acknowledged that AI-use falls along a spectrum, from benign to problematic, and shared the subjective factors that they considered when judging AI use.
Some participants spoke of the \textit{intention} of a poster\footnote{We use the term \emph{poster} to refer to both those posting top-level subreddit submissions as well as those posting comments in response to submissions.}, like a moderator of r/itookapicture\footnote{To provide additional context, we introduce quotes with the name of the subreddit that the participant moderates. In the case where the participant moderates several subreddits, we include either the subreddit that the quote is referring to or, for more general quotes, the largest subreddit that the participant moderates.} who said: \emph{``I don't mind people using AI to better state their thoughts nor create art as long as it is not being used deceptively.''}
Others referenced the \textit{degree} to which a poster relied on AI, like a moderator of r/ChangeMyView who reported being OK with a post made with AI, \emph{``if it was sufficiently [a poster's own] words, just sort of juiced up by ChatGPT.''}
Another participant expressed similar concern that posts should be a poster's \emph{``own intellectual contribution or content''}:

\begin{interviewquote}
\textbf{r/AskHistorians moderator:} So say, for example, we have an expert\ldots perhaps they're an expert on German history, but they don't speak English all that well. So they write their answer in German, and then use ChatGPT to try to translate it\ldots it is their own intellectual contribution or content, they've created this themselves, but then they just use this as a tool in a perfectly viable way.
\end{interviewquote}

Other participants took a less nuanced view, like a moderator of r/WritingPrompts, who relayed their community's official stance: \emph{``Let's be absolutely clear: you are not allowed to use AI in this subreddit, you will be banned''}.
Despite hearing arguments that, \emph{``ChatGPT is another writing tool that authors are using and banning ChatGPT will be the same as banning a spellchecker,''} this participant disagreed and categorically rejected all uses of AI, going so far as to say that, \emph{``[AIGC] is antithetical to everything that [our] subreddit is about.''}
The simplicity of this stance has its appeal, as more subjective definitions of AIGC introduce the possibility that community members will disagree about specific cases---an implication which we will revisit in later sections.

\subsection{Findings: Moderator Concerns About AIGC (RQ1)}\label{objections}
Our recruitment targeted moderators of communities that are actively restricting the use of AIGC. 
Accordingly, while our participants offered differing definitions of AIGC, they generally opposed its use in their communities.
We group our participants' reported AIGC concerns into three, often overlapping, categories, summarized in Table~\ref{tab:objections}. 
Content quality concerns (detailed below in Section~\ref{content-objections}) take issue with the quality of AIGC and worry that it decreases a community's utility.
Social dynamics concerns (Section~\ref{social-dynamics}) focus on how AIGC reduces the social value that a community provides its members. 
Governance concerns (Section~\ref{governance}) emphasize the ways in which AIGC makes the job of moderating online communities more difficult. 
Naturally, these categories overlap and interact, and participants often described considerations that span multiple.
Despite general opposition, participants also acknowledged potential benefits of AIGC (Section~\ref{benefits}).

\begin{table}
\small
\caption{Summary of our participants' main concerns about AIGC in their communities.}
\begin{tabular}{|c|c|}
\hline
\textbf{Concern} & \textbf{Examples}                               \\ \hline
Low Quality Content            & Poor prose, factual inaccuracy, off topic \\ \hline
Impacted Social Dynamics              & Fewer authentic connections, strained relationships, ideological opposition \\ \hline
Difficult to Govern            & Increased volume of community attacks, harder to detect deception \\ \hline

\end{tabular}
\label{tab:objections}
\end{table}

Moderators shared various perspectives on the magnitude of these concerns.
When asked if AIGC was one of their top concerns, a moderator for r/CryptoCurrency noted that, \emph{``Early 2023, I would've said yes, but right now it has calmed down quite a bit.''} 
On the other hand, some participants did regard AIGC as a major concern:

\begin{interviewquote}
\textbf{r/news moderator:} The concern is the more weaponized stuff\ldots the folks who are utilizing these tools for very focused purposes, often for political reasons, or for commercial reasons, those are the ones that I think concern us the most. Whether or not somebody manages to get a post up there that doesn't belong for a few hours, most of us don't care\ldots But what we do care about is creating a veneer of legitimacy for bad actors.
\end{interviewquote}

We will explore the challenges of detecting AIGC in Section~\ref{detecting}, but first we dig deeper into the types of concerns that our participants shared.

\subsubsection{Content Quality Concerns}\label{content-objections}
First and foremost, moderators were concerned about the quality of AIGC.
Nine participants reported that AIGC did not meet the quality standards of their subreddits, like a moderator of r/AskHistorians who said that AIGC, \emph{``tries to meet the substance and depth of a typical post\ldots however, there are frequent glaring errors in both style and content.''}
Participants shared several reasons that they considered AIGC low quality, such as issues with its style, its tendency to be factually inaccurate, and the belief that it was off-topic.

Several participants reported that certain stylistic issues with AIGC make it low-quality content.
One moderator said that AI-generated answers to questions often, \textit{``do not address the question being asked\ldots tend to be very general and `hedge' more than a real human.''}
Other moderators put it more bluntly, like a moderator of r/CryptoCurrency who noted that, \emph{``[AIGC] is not enjoyable to read''}, or a moderator of r/Eldenring who told us: \emph{``[AIGC] is content that provides no value/discussion at all, it's low effort content basically.''}
This moderator considered effort to be a signal of value or quality, which we will revisit in the discussion.
They went on to share that this sort of low effort content can, \textit{``[discourage] users who want to post their own original content''}.

Four participants who moderate communities that emphasize the accuracy of information (e.g., r/AskHistorians and r/explainlikeimfive) specifically mentioned their concern about inaccurate AI hallucinations.
This concern is captured by a moderator of r/explainlikeimfive who objected to AIGC on the grounds that ChatGPT was a \emph{``bullshit generator''}, or by another participant who shared that AIGC's inaccuracy was unacceptable:

\begin{interviewquote}
\textbf{r/AskHistorians moderator:} History and the work thereof is something that I'd say AI will continue to find impossible for the foreseeable future, until such time as it can gain a sense of truth, or at least coherence.
\end{interviewquote}

This participant described AIGC as \emph{``truthy-seeming''} because of its tendency to include fake citations and felt that the appearance of being thoroughly researched increased the chances that readers would believe it, which, they feared, would \textit{``undermine the trust that people have [in the community].''}

Finally, several moderators said that AIGC was simply off-topic in their communities.
For example, a moderator of r/itookapicture shared that the goal of their community was, \textit{``sharing great photography and photography techniques,''} and accordingly, \textit{``preventing the submission of AI-generated images and human-created compositions alike are a top priority.''}

\subsubsection{Social Dynamics Concerns}\label{social-dynamics}
Of course, online communities are not only a source of high-quality content but also of meaningful social interactions.
Ten participants said that they were concerned that AIGC would negatively impact the social dynamics in their communities, reducing the social value that their communities provide to members.
Participants mentioned several ways that this could happen: decreasing opportunities for human connection, straining relationships, and violating their communities' values.

Ten participants reported that AI went against their communities' purpose, which was to share posts created by humans.
These participants mentioned the inherently inhuman nature of AIGC and expressed concern that this would lead to fewer authentic connections in their communities.
These were not necessarily objections to the quality of AIGC, but rather to the \emph{process} through which it is created:
\begin{interviewquote}
\textbf{r/explainlikeimfive moderator:}
In our case, we do want someone to write these explanations. And we don't want it done by some AI that has no clue what it's talking about. And even if it does give an accurate answer, the purpose of this site is for people to write in their own words \ldots there's already Google, there's already Wikipedia.
\end{interviewquote}
These participants believed people come to their community to connect with others through dialogue and that AIGC does not meaningfully contribute to---or even detracts from---such conversations.
A moderator of r/changemyview summarized this sentiment well: \emph{``How can we change your view when you haven't actually stated your view at all? This is ChatGPT’s view. And if I wanted to, I could go argue with ChatGPT. But that's not helping here.''}
Six moderators of communities of practice (e.g r/itookapicture, which is devoted to photography) similarly said that AIGC did not contribute to the collaborative learning that the community aimed to encourage.
These are communities where members seek to develop particular skills by doing, sharing, and receiving feedback from other practitioners.
Accordingly, posts produced \textit{without using} their practice did not provide educational value to the community.
A moderator of r/WritingPrompts summarized this concern when they shared that their community existed to, \emph{``[give] writers a chance to practice their craft and practice their skill. And if you are taking away the practice element of that, because all you're doing is feeding a prompt into ChatGPT\ldots The server is no longer serving its purpose.''}

Additionally, participants shared concern that AIGC could strain relationships in their community.
For example, a moderator of a political subreddit noted that: \emph{``the possibility [of AIGC] fuels Redditors accusing other comments of being written by AI (which is a form of incivility).''} 
Beyond incivility, the controversial nature of AIGC could cause discussions to devolve into off-topic arguments:
\begin{interviewquote}
\textbf{r/Zelda moderator:} Many comment sections would fill with the same discussion points about the controversies of AI art\ldots We would often see comments that attacked people for using AIGC, as well as attacks on people for gatekeeping art. 
\end{interviewquote}

Finally, some participants took an ideological stance and objected to the use of AI and, by extension, AIGC, because they saw it as incompatible with their communities' values.
For example, a moderator of a creative subreddit objected specifically to the way that generative AI models are trained:

\begin{interviewquote}
\textbf{r/<removed>\footnote{We use \emph{<removed>} to disassociate subreddits from participant quotes either at the participant's request or when we find the quote together with the subreddit name potentially too revealing.} moderator:} Perhaps most importantly, as stewards of a creative space, we feel it's our duty to support the real human artists, authors and other creatives whose work has been exploited to train the vast majority of these models without their knowledge and without credit or compensation. 
\end{interviewquote}

\subsubsection{Governance Concerns}\label{governance}

Participants also objected to AIGC because it made the job of moderating their communities more difficult.
Eight participants said that they saw malicious uses of AIGC and were concerned that existing moderation techniques would struggle to handle it.
These moderators spoke of pre-existing problems, such as spam and harassment, that are more difficult to manage when they are amplified in scale or sophistication by AIGC. 
While AIGC may be less common than other moderation challenges, it can still be quite disruptive:

\begin{interviewquote}
\textbf{r/explainlikeimfive moderator:}
It's not our most common removal, by far. But personally, I would rate it as the most threatening concern\ldots It's often hard to detect and we do see it as very disruptive to the actual running of the site.
\end{interviewquote}

Five participants, who moderated the largest subreddits, spoke of AIGC being used in broad attacks on their communities.
These attacks often involved automated accounts that post large amounts of spam to disrupt community functioning.
One moderator discussed having to ask Reddit staff for help:
\begin{interviewquote}
\textbf{r/AskHistorians moderator:}
A few months ago, we had an instance where we were subject to a bot attack that was using ChatGPT\ldots we ended up reporting it to the admins who were able to take care of it through whatever they do on their back end.
\end{interviewquote}
Other times these attacks aim to deceive users as a part of influence operations.
A moderator of r/news, one of the platform's largest subreddits, shared a story of one such operation, which they described as, \emph{``a highly sophisticated deployment, an attempt to sway a narrative by utilizing accounts that had had their karma bolstered by automated commenting.''}
This participant references Reddit \textit{karma}\footnote{\url{https://support.reddithelp.com/hc/en-us/articles/204511829-What-is-karma}}, a reputation points system that communities can use to limit who can post, noting that AIGC can be used to bypass this system through a practice known as \emph{karma-farming}, in which automated accounts make frequent posts in order to rapidly accumulate karma.
Karma-farming is not a new practice, but this moderator thought that AIGC made it harder to detect, and thus more effective.
Beyond karma-farming, we also heard of other forms of deception that used AIGC:

\begin{interviewquote}
\textbf{r/CryptoCurrency moderator:} We’ve seen some `shills' (aka companies or group of people) coming with AI content to promote their product\ldots And of course promoting is detrimental, as it is often disguised as a `research' post about that specific coin (for example), making it seem like a legit user researching a subject and that might influence buyers in comparison to a regular ad.
\end{interviewquote}

In addition to broad attacks on a community, we also heard from participants who were concerned about AIGC being used for targeted attacks on individual members. 
One moderator of r/politics reported that they associated AIGC with, \emph{``usually some form of trolling, or `outsourcing' writing a comment the user wanted to make.''}
Another participant also mentioned trolling, though they said that existing moderation solutions for this problem would be sufficient:

\begin{interviewquote}
\textbf{r/changemyview moderator}: The one thing that I am worried about AI is that it does make trolling easier. You can generate larger amounts of trolling text and make it seem reasonable or seem like somebody wrote it for a lot longer\ldots But\ldots it's not that hard to tell when somebody's trolling on our subreddit anyway. So it makes trolling easier [to do], but it's something that I don't think we're going to be too stressed handling.
\end{interviewquote}

Finally, participants expressed concerns about AI making it easier to post objectionable content.
One participant who moderated r/Zelda shared that they had seen AI used to produce specific types of objectionable content that their community had already banned, such as commercial or NSFW content.
They noted that while this type of content was not new, \textit{``it seemed to be new territory for people to expedite these things via AI.''}

\subsubsection{Potential Benefits of AIGC}\label{benefits}
While our recruiting approach and research questions primarily surfaced an overall attitude of opposition, several moderators mentioned ways that AIGC could potentially \textit{benefit} their communities by increasing engagement. 
These moderators perceived a legitimate, or at least good-faith, motivation behind posters who used AIGC to complement their knowledge and writing ability:
\begin{interviewquote}
\textbf{r/AskHistorians moderator:} We're a very popular subreddit, and posts tend to be held up as examples of excellent content on Reddit. Some people may want to be a part of it, but not have the specific knowledge or resources to be able to contribute `properly'. Hence they turn to AI tools, hoping to provide a decent answer. 
\end{interviewquote}

Three moderators mentioned that AI could help non-native English speakers improve their posts.
A moderator of r/changemyview reported that, \emph{``AI as a tool is going to be useful\ldots it's useful for people who are poor writers, useful for people who aren't strong English speakers. And when they need help, AI can fill in those gaps,''} though they noted that it was important that such a contribution was still, \emph{``sufficiently their words''}.
Of course, the definition of ``sufficient'' will be context dependent and may change over time as norms around AI use evolve.

One last benefit shared by our participants was that AIGC could increase engagement in an inactive conversation.
For example, one moderator speculated about the potential benefits of AIGC for platforms and provided a hypothetical example:
    
\begin{interviewquote}
\textbf{r/<removed> moderator:} So a subreddit that needs a little bit of life breathed into it, [Reddit] deploys some AI there and make a variety of posts. And then people comment and they see a whole bunch of ads. And everyone's happy, right? A weekly automated thread isn't necessarily a bad thing.
\end{interviewquote}
This moderator was not saying that their community had done this or that it aligned with their values, but rather that they believed this approach could be useful in certain contexts.

On the whole, our evaluation of the interview transcripts suggests that despite these potential benefits, all of our participants felt that, at least for now, the negative aspects of AIGC outweigh the good.

\subsection{Findings: Online Communities' Responses to AIGC (RQ2)}\label{responses}
To counter the perceived threats of AIGC, moderators have enacted specific changes in their communities. 
First, our participants emphasized collaborating with their community members to clarify norms about the use of AIGC (Section~\ref{norms} below).
Once clarified, these norms could be codified into rules, which moderators were then tasked with enforcing (Section~\ref{enforcing}).

\subsubsection{Clarifying Community Norms}\label{norms}
Most participants spoke of establishing a community stance towards AIGC through a collaborative and iterative process involving community members and moderators.
Often it was the moderators who started the discussion, but other times it was the community:

\begin{interviewquote}
\textbf{r/museum moderator:} It was the users who protested. Mildly: it's a fairly polite subreddit. They said, we don't believe that this belongs here. And so we had a blanket ban on it for a little while. And then we slowly allowed it to be integrated. And then there was a counter protest [against the integration].
\end{interviewquote}

Most participants reported they wanted to ensure that their stance on AIGC was aligned with their communities' interests.
One participant said that their community initially allowed AIGC before eventually voting to ban it:
\begin{interviewquote}
\textbf{r/CryptoCurrency moderator:} At first (late 2022 early 2023) we let people use AI to write partial content in their posts *if they mentioned it*. But lots of abuse, people not mentioning it, and just overall quality decreasing, so [the community] decided to ``ban'' AI use.
\end{interviewquote}
This participant went on to describe their community's process for discussing issues, surfacing proposals for solutions, and soliciting proposal votes from all community members---a process that they used here to enact a rule banning AIGC.

Almost all participants mentioned eventually codifying community norms into public subreddit rules.
Some communities used rules that explicitly banned AIGC or bots, while others used more general rules (such as rules against plagiarism) as effective AIGC bans.
Participants shared mixed results about the efficacy of these rules.
Seven participants said that the volume of AIGC had decreased in response to new rules, including a moderator of r/AskHistorians who said, \emph{``People are using [AIGC] a little bit less now that they know what the rules are. They know that people don't necessarily want it.''}
However, a moderator of r/explainlikeimfive expressed skepticism about the efficacy of rules: \emph{``I think it may stop some of the average users (to the extent that they actually read the rules, which is a reddit-wide problem.) Bots obviously don’t read rules.''} 
As this participant suggested, explicit rules can guide the behavior of good-faith users, but might not deter bad actors.

\subsubsection{Enforcing Community Rules}\label{enforcing}
All of our participants' communities devised---and tried to enforce---rules about AIGC.
However, the difficulty of detecting AIGC meant that enforcement of such rules was often challenging.  
Three moderators spoke of AIGC increasing their workload:
\begin{interviewquote}
\textbf{r/AskHistorians moderator:} It's not fun to deal with having to assess whether or not something is [AIGC]. Reading through it like that, it's a lot of extra work, right? Because you have to read through it all the way. It's so fast for people to input this kind of content and a lot slower for people to assess it.
\end{interviewquote}
This imbalance between the effort to produce AIGC and to read it, given AIGC tends to be verbose, puts moderators at a disadvantage.
Also, given how alienating a false accusation can be, moderators must take extra care with their decisions; a moderator of r/explainlikeimfive shared, \emph{``You don't want to be banning someone for using GPT when they don't actually use it. So we might have to watch someone for a bit of time.''} 

Our participants used a variety of strategies to make enforcement of these rules less burdensome.
Seven moderators mentioned sharing the work with community members, like a moderator of r/news who noted, \emph{``The first line of defense is the users themselves.''}
However, as mentioned in Section~\ref{social-dynamics}, this runs the risk of spurring false accusations and incivility between members.

Four moderators spoke of their reliance on platform tools, especially \emph{automoderator}\footnote{\url{https://www.reddit.com/wiki/automoderator/}}, a programmable bot that scans and performs actions on all new posts.
Such automated tools lessened the manual effort required by moderators, but they are only as effective as the detection heuristics programmed into them, which, as we discuss in the next section, can be flawed. 
Moderators said that such tools are not enough to catch all AIGC, and in cases of extreme post volume, such as during coordinated bot attacks, moderators sometimes sought help from Reddit platform admins:

\begin{interviewquote}
\textbf{r/news moderator:} There’s the stuff that's publicly available to every user, then you have the stuff that's available to Moderator accounts, within the subreddit that the moderators are working on, and then you have the admin stuff---which is much more sophisticated than what we have.
\end{interviewquote}

All of these enforcement techniques ultimately require making decisions about what is and is not AIGC. 
Next we discuss our participants' reported methods for making these difficult decisions.

\subsection{Findings: Methods for Detecting AIGC (RQ2)}\label{detecting}
Our participants shared custom heuristics (Section~\ref{heuristics}) and technical solutions (Section~\ref{tools}) that they use to detect AIGC. 
Even if participants thought their detection techniques mostly work now, they worry that their tools will be less effective as generative AI improves:

\begin{interviewquote}
\textbf{r/news moderator:} We see the obvious stuff and we prune the obvious stuff\ldots but there has to be a lot that we're missing. And I imagine that it's gonna get more and more sophisticated over time.
\end{interviewquote}

\subsubsection{Detection Heuristics}\label{heuristics}
There are a number of signals that moderators believe help them identify AIGC. 
To avoid aiding bad-faith actors, we do not list specific tells, but they broadly fell into the following categories: recognizable language patterns in posted content, details of users' accounts and behavior, deviation from a community's style norms, and inaccurate information.

\paragraph{Content Signals}
Twelve participants mentioned that certain patterns in a post's content, such as keywords, phrases, or distinct forms, may make moderators suspect that a post is AIGC.
A moderator of r/Eldenring noted that AI-generated text is, \emph{``very repetitive i.e it tries to justify a point but instead of doing that, it ends up repeating the same point over and over to the point of ad nauseum.''} 
A moderator of r/itookapicture shared that with AI-generated images, \emph{``there is often something that feels a bit unnatural\ldots
The images look like the average of something, rather than a unique individual with all of the flaws that come with that.''}

\paragraph{User Signals}
When confronted with suspect content, six participants reported that they often consider the details of a poster’s account, including their past posting behavior.
One red flag is a dramatic change in the style of a poster's prose. For example, a moderator of r/changemyview shared that suspicious posts were,
\emph{``often very different in terms of writing from what the user posted in response to other users questions or comments.''}
Additionally, four moderators, like a moderator of r/Zelda, noted that signals that identify more traditional bots are also relevant for identifying AIGC: \emph{``accounts used for spam\ldots there are other indicators to find those (account age, history,\ldots) Even before ChatGPT, we would find copy-pasted or markov-generated comments with these same account indicators.''}

\paragraph{Deviation From Subbreddit Style Norms}
Four participants reported using their knowledge of their communities’ particular communication norms to identify atypical posts as AIGC:

\begin{interviewquote}
\textbf{r/AskHistorians moderator:} It feels very formal, and it feels very different from the normal kinds of comments. There's definitely sort of a genre to answering a question on r/AskHistorians that people who have been around for a while tend to all follow, just because there are certain ways to communicate historical knowledge in an open online space to a group of non experts.
\end{interviewquote}

\paragraph{Inaccurate Information}
Four participants reported using their domain knowledge to identify posts as false or incoherent, which to them suggests AIGC:

\begin{interviewquote}
\textbf{r/museum moderator:} In r/museum\ldots there is a Canon, so to speak, of material. So it's not like there's suddenly going to be a brand new, you know, 16th century French academic painter that nobody's ever heard of. And so something that seems incongruent to that situation would obviously raise an eyebrow.
\end{interviewquote}

Our participants reported that these heuristics worked, but also described them as time-consuming. 
Participants thus turned to technical tools to speed up the detection process.

\subsubsection{Technical Tools}\label{tools}
Though they acknowledged such tools' unreliability, most participants used automated approaches to detect AIGC, including both third-party detection tools and community-developed solutions.
Participants spoke of programming the detection heuristics mentioned above into tools such as \emph{automoderator}, which apply the heuristics to every new post and flag anything suspect.
Six participants mentioned manually copying posts into third-party detection tools to see if they were AIGC. 
A moderator of r/writingprompts reported that they had tested, but never deployed, a homespun bot that would apply third-party detectors automatically to all new posts.

At the same time, a moderator of r/changemyview mentioned that because they are aware of the unreliability of third-party detection tools, they rarely defer to them absolutely: \emph{``I don't necessarily trust the tool--that's why we have several layers.\ldots
We always try to talk to the user.''}
Two participants mentioned that the tools were least effective on shorter text, like a moderator of r/writingprompts who noted that one tool, \emph{``was very bad with short comments. If a comment was like 100 words or a really short story--every poem got flagged as AI. And like, No, it's not. It's just short.''}

\section{Discussion}
Our qualitative investigation offers two main research contributions: (1) a description of moderators' attitudes towards AIGC in subreddits that restrict its use and (2) insights into how these moderators and their communities have responded to the early challenges of AIGC. 
On the whole, our findings align with those of recent studies into the impact of generative AI on online communities~\cite{burtch_consequences_2024,wei2024understandingimpactaigenerated}, which emphasize that the \textit{social value} of online communities is more important than ever in the age of generative AI. 
As a company, Reddit seems to agree: in a May 2025 update to the community~\cite{spezRedditsNextChapter2025} Reddit's CEO shared a similar sentiment:
\begin{interviewquote}
Reddit works because it’s human. It’s one of the few places online where real people share real opinions. That authenticity is what gives Reddit its value. If we lose trust in that, we lose what makes Reddit\ldots Reddit. Our focus is, and always will be, on keeping Reddit a trusted place for human conversation.
\end{interviewquote}
Platform designers and community stewards should both focus on ways to increase this social value, and one way to do this is by empowering communities to take their own stances on AIGC.

As generative AI usage grows, we place our work in dialog with Seering's call to the CSCW research community to center community self-moderation~\cite{Seering20}, especially their second ``guiding question'': \emph{``What are the processes of context-sensitivity in online community moderation, and how might they be better supported?''}
We believe our findings demonstrate that allowing communities to make context-sensitive decisions about AIGC is central to their self-determination and should be an important topic for CSCW research.
Unwanted AIGC in online communities is an active and growing~\cite{lloydAIRulesCharacterizing2025} concern, with Reddit's CEO describing it as \textit{``the worry I hear most often these days from users and mods alike.''}~\cite{spezRedditsNextChapter2025}
To offer design implications that may help communities address this concern, we expand below on the values that shape community attitudes towards AIGC as well the importance of clear and enforceable community norms.

\subsection{Community Values Shape Attitudes Towards AIGC}
In a recent quantitative study~\cite{lloydAIRulesCharacterizing2025}, we demonstrated that rules about AI are most common in certain types of subreddits, like those that are larger and those devoted to Content Generation.
Our current study digs deeper into the motivations behind these rules and reveals a spectrum of moderator attitudes that vary with the purpose and size of communities.
We expect some objections to AIGC will lessen over time, while others will be more persistent.

\subsubsection{Community Types and AIGC Attitudes}
Our participants' concerns about AIGC were largely based on the \textit{types} of communities they moderated.
Accordingly, it is useful to consider our results through the lens of several online community types from the Social Computing literature, such as ~\emph{knowledge communities}~\cite{jenkins_introduction_2006}, \emph{empathic communities}~\cite{preece_empathic_1999}, ~\emph{communities of practice}~\cite{lave1991situated}, and the five \textit{Community Archetypes} presented by Prinster et al.~\cite{Prinster24}: Topical Q\&A, Learning \& Perspective Broadening, Social Support, Content Generation, and Affiliation with an Entity.

We heard much about AIGC's threat to the social value of a community (Section~\ref{social-dynamics}), which we suspect is most concerning to communities that emphasize authenticity~\cite{Smith23}, like empathic communities~\cite{preece_empathic_1999}, or Social Support communities.
Interestingly, our prior quantitative study~\cite{lloydAIRulesCharacterizing2025} found that Social Support communities are \textit{negatively} associated with having a rule about AI.
Explicit rules may be less likely in these communities, but our qualitative findings suggest that this should not be interpreted as a lack of concern.
Participants who moderated~\emph{knowledge communities}~\cite{jenkins_introduction_2006} like r/AskHistorians and r/explainlikeimfive, which fit the Topical Q\&A archetype, also shared their concerns about AIGC's impact on the social value of their communities.
However, these participants additionally discussed concerns about the inaccuracy of AIGC (Section \ref{content-objections}). 
Arguably the clearest objections came from moderators of \emph{communities of practice}~\cite{lave1991situated} like r/itookapicture and r/WritingPrompts, which fit the Content Generation archetype. 
These communities shared content quality and social dynamics concerns about AIGC, which have additionally been documented in other recent work~\cite{matatovExaminingPrevalenceDynamics2025}.
While long-standing communities of practice may struggle to adapt to AI, we expect that others will emerge specifically devoted to AI use\footnote{\url{https://www.reddit.com/r/aiArt/}}.
Moderators of larger, more impersonal communities that fit the Learning \& Perspective Broadening archetype, such as r/news, r/CryptoCurrency, and r/politics, primarily voiced governance concerns (Section ~\ref{governance}). 
We suspect this is because the size of these communities make them a target of bad actors.
Finally, we primarily heard ideological objections from the communities in our sample that fit the Affiliation with an Entity archetype.
We suspect this is because in our sample these are Content Generation communities (r/WoW, r/Zelda), and it remains to be seen if communities affiliated with non-media entities, like geographical places or organizations, would share similar concerns.

On the whole, our participants rejected AIGC because they felt the main purpose of their communities was to share posts created by humans---this was true across all of the community types that we encountered (of course, remember that our sampling strategy focused on communities with concerns, not representativeness).
This sentiment was made clear by the recent research ethics controversy on r/changemyview, where community members were outraged to learn that researchers had been posting undisclosed AIGC~\cite{changemyviewMETAUnauthorizedExperiment2025}.

\subsubsection{Attitudes May Shift Over Time}
Our findings indicate that community norms about AIGC are still in flux.
Perhaps over time people will come to treat generative AI in the same way they treat tools like auto-complete or spellcheck (a possibility we discussed in Section~\ref{definitions}).
If so, process-based objections may fade away.
Of course, the opposite could happen as well: as generative AI becomes more pervasive people may seek out communities of collective refusal. 
In order to support community self-determination, we should design platforms that allow for both possibilities.

Additionally, generative AI will continue to change, and in time the content-based objections that our participants shared may no longer be relevant.
For example, we consider objections based on AIGC's accuracy and prose quality, as well as those based on unethical model training processes, to be potentially addressable by technological advances. 
If online communities are to remain relevant, moderators will need to stay abreast of their members' stance on this evolving technology;
ultimately, if moderator and member perspectives on AIGC differ too much, Reddit's distributed moderation system allows users to ``agree to disagree" by leaving and finding other, more aligned communities~\cite{grimmelmann2015virtues}.
At the time of writing, some platforms have begun incorporating generative AI into their products~\cite{lunden_linkedin_2023}, likely in an effort to increase engagement.
Though this integration might be aligned with platforms' profit motives, our interviews suggest it is not aligned with the preferences of all users or communities, as the low effort required to use these features will likely increase the prevalence of AIGC.
As we mention in Section~\ref{benefits}, generative AI might bring certain benefits, but given the strong objections that we have seen, attempts by platforms to integrate generative AI should be done via a collaborative design process involving users~\cite{Matias18}.
We suggest that platforms that host a variety of community types, such as Reddit, give individual communities the choice to opt-out of generative AI features.
If communities choose to opt-out, this decision could be communicated to posters via the UI, perhaps by displaying the tools in a disabled state with an informative tooltip.
Signaling a community's stance on AIGC in this way can have the added benefit of discouraging the use of AIGC that is produced off-platform.
Additionally, for platforms designed for just one of the community types that we discussed, we recommend considering the community type's values when deciding whether to add generative AI features. 
Otherwise, platforms run the risk of alienating users~\cite{Smith23}.

\subsection{The Challenges of Establishing and Enforcing Norms}
Our findings relate to and reflect on research about the power of norms to shape behavior in online communities~\cite{chancellor18,Chandrasekharan18,kiesler11,Lessig99}.
We explore the specific challenges of establishing and enforcing community norms about AIGC and discuss ways in which platform designers might be able to aid community moderators with these tasks.

\subsubsection{Clarifying Norms in a Moment of Flux}
Our interviews show that self-governing online communities are adapting to AIGC by clarifying norms and codifying these into rules.
Prior research has shown that online communities often make rules in response to sudden changes~\cite{SeeringEtal2019mec} and that this rule-making can be an effective technique for navigating transitions~\cite{Kiene16}. 
Moderators sometimes struggle to identify AIGC, which makes enforcing rules about it a challenge; even so, discussing and clarifying rules can help inform the behavior of well-intentioned community members. 

As noted above, community norms around AIGC are in flux and likely will be for some time.
Community members and moderators may have different understandings of whether or not AIGC is acceptable, and this misalignment~\cite{Koshy23} can result in well-intentioned users inadvertently violating community norms.
Communities can reduce misalignment by making their norms more explicit and known to members, which has proven effective with other moderation challenges~\cite{matias2019preventing}.
Indeed, our participants' report that AIGC has decreased since they enacted bans, though these trends have not been independently verified and may change over time.
Platform designers can help by updating UIs to ensure that well-meaning posters are aware of community rules. 
For example, platforms could update their authoring UIs to display a community's rules, perhaps with a checkbox that users must mark to indicate that they have read and will abide by the rules, similar to how privacy policies are commonly presented.
Another option would be to display the rules as placeholder text in the UI element where posts are written, which may remind users of norms at the time of posting~\cite{Kiene16}. 
That said, displaying rules in a creative way may not be enough, as past research has demonstrated that privacy policies are rarely read~\cite{obar_biggest_2020} and many posters simply do not read a community's rules before posting~\cite{JhaverAppling19}.
Communities may also consider more thorough ``onboarding processes'' to ensure new members are familiar with community rules, such as an interview with an existing member. 
However, this approach adds an additional time burden and might deter potential members who are wary of such a commitment.

\subsubsection{Equitably Enforcing AIGC Bans}
Our findings detail some of the challenges that moderators currently experience enforcing AIGC restrictions and bans.
Even before the AIGC challenge, moderators were stretched thin trying to protect their communities from problematic content (e.g., mis- and disinformation) and behavior (e.g., harassment, incivility, and influence campaigns). 
With detection of AIGC being an unsolved problem~\cite{edwards23}, moderators rely on heuristics to enforce restrictions or bans. 
Our findings show that these heuristics can be time-intensive. 
Even worse, they may be biased, as other work has shown that people's intuitions about AIGC are wrong in predictable ways~\cite{Jakesch_2023} and that certain social groups are more often suspected of AI use~\cite{kadoma2024generativeaiperceptualharms}.
Such consistent bias could disproportionately affect minorities and individuals seeking legitimate participation~\cite{liang23}, such as non-native speakers, newcomers, and gender or racial minorities. 
It is concerning that most of our participants were confident in the efficacy of their heuristics and unaware of any potential bias.
Our participants' reliance on these heuristics suggests that moderators should be educated about potential bias and that further studies should evaluate the efficacy of detection heuristics in content moderation.
Platforms may be able to help moderators by adding tools for AIGC detection.
Earlier work demonstrated that ``synthesized social signals,'' or ``social signals computationally derived from an account's history,'' help community members evaluate the credibility of other accounts~\cite{Im20}.
Similar tools may be useful for detecting AIGC.
For example, showing information about a poster's interactions with the system, such as time spent editing a post, could be helpful, though this also has user privacy implications.
Surfacing a poster's past behavior and other community memberships could also be relevant, as these have been shown to be useful in identifying misinformation posters~\cite{Bozarth2023}.

Given the combination of concerns about AIGC and the lack of tools to detect it, we expect to see communities develop their own, non-technical, means for demonstrating authenticity.
Communities may have no choice: they may struggle to function, as recent work has suggested that AI makes cooperation harder in pseudonymous, low-trust settings like online communities~\cite{wojtowiczUnderminingMentalProof2025}.
Past research provides examples where novel communication practices have emerged in anonymous online communities, such as \emph{triforcing} (posting complex Unicode symbols only meaningful to the community) and \textit{timestamping} (posting photos of oneself holding a piece of paper with the current date and time on it)~\cite{bernstein14}.
These techniques signal in-group membership and authenticity by presenting something that only an authentic poster can produce.
These techniques also demonstrate effort, which---as a participant mentioned in Section~\ref{content-objections}---makes a contribution seem more valuable.
Past research calls this type of communication \textit{effortful communication} and documents how it can facilitate connection between online community members~\cite{Zhang22}.
Certain interface designs can encourage such communication, but
it remains to be seen if these techniques can offset the negative effects that AIGC may have on interpersonal communication.
In the age of low-effort AIGC we expect to see online communities place an even greater premium on effortful communication, which platform design can encourage.
While these social innovations may emerge naturally from within a community, platforms can provide technical affordances to support their adoption, and the HCI research community can help.

\section{Limitations and Future Work}
Our qualitative study is based on interviews with moderators of communities that were taking steps to govern the use of AIGC.
This recruiting criteria supported our goal of understanding some of the attitudes and practices of Reddit moderators as they adapt to the growing prevalence of AIGC.
Our recruitment criteria and limited sample means that our study and results do not necessarily explain how \textit{all} Reddit moderators and communities consider AIGC.
Fully understanding the perceptions, attitudes, and impact of generative AI on online communities requires future work that approaches this topic from different angles.

For example, our study purposefully omitted communities that have fully embraced AIGC, as well as those without AI-specific rules.
Studying communities who encourage the use of AIGC could be a fruitful direction for future work.
Our recruitment additionally focused on larger subreddits in order to target those that are currently most affected by AIGC. 
While our participants also moderated smaller subreddits and shared how their experiences and concerns are relevant to communities of various sizes, additional studies could include moderators from a wider set of communities.
Such studies could provide a more complete picture of the range of ways that online communities are responding to AIGC, and to further explore the relationship between a community’s values~\cite{Prinster24} and its stance on AIGC.
For example, the AI concerns that we identified may be specific to certain subreddit topics, and our limited sample may have missed other important concerns.
A recent study of ours used computational content analysis to understand how these qualitative findings generalize across subreddits of different size and type~\cite{lloydAIRulesCharacterizing2025}. 
Other work could use surveys or other quantitative methods to more fully understand the impact on communities.
Further, since AIGC is an internet-scale phenomenon, future studies should look beyond Reddit in order to understand how moderators are responding on platforms with different purposes and norms.

\section{Conclusion}
We performed a qualitative investigation into volunteer Reddit moderators' experiences with, and attitudes towards, AI-generated content.
Moderators shared that they are concerned about AIGC impacting the content quality, social dynamics, and governance processes in their communities, despite also seeing its potential to make it easier for some community members to contribute. 
In response, communities have enacted bans, which moderators enforce using time-intensive and imperfect heuristics.
Despite enforcement challenges, our participants reported that their rules are currently sufficient, as they clarify community norms about AIGC use.

As generative AI usage grows, the CSCW community needs to pay attention to how it affects the dynamics, practices, and health of online communities. 
There is much that the research community can do to develop tools and practices that may help community moderators guide their communities through this period of technological change.
As a starting point, this paper documents the perspectives of moderators at this critical juncture, where, as one moderator commented, ``there has to be a lot that we're missing.''

\begin{acks}
This material is based upon work partially supported by the U.S. National Science Foundation under Grants No. IIS 1901151, SaTC: 2120651, and No. 2313998.
Any opinions, findings, and conclusions or recommendations expressed in this material are those of the author(s) and do not necessarily reflect the views of the U.S. National Science Foundation. 
Thank you to Sarah Gilbert for valuable feedback at various points in this study. 
\end{acks}

\bibliographystyle{ACM-Reference-Format}
\bibliography{sample-base}


\begin{thebibliography}{99}


\ifx \showCODEN    \undefined \def \showCODEN     #1{\unskip}     \fi
\ifx \showISBNx    \undefined \def \showISBNx     #1{\unskip}     \fi
\ifx \showISBNxiii \undefined \def \showISBNxiii  #1{\unskip}     \fi
\ifx \showISSN     \undefined \def \showISSN      #1{\unskip}     \fi
\ifx \showLCCN     \undefined \def \showLCCN      #1{\unskip}     \fi
\ifx \shownote     \undefined \def \shownote      #1{#1}          \fi
\ifx \showarticletitle \undefined \def \showarticletitle #1{#1}   \fi
\ifx \showURL      \undefined \def \showURL       {\relax}        \fi
\providecommand\bibfield[2]{#2}
\providecommand\bibinfo[2]{#2}
\providecommand\natexlab[1]{#1}
\providecommand\showeprint[2][]{arXiv:#2}

\bibitem[AI(2022)]%
        {OpenAI22}
\bibfield{author}{\bibinfo{person}{Open AI}.} \bibinfo{year}{2022}\natexlab{}.
\newblock \showarticletitle{Introducing ChatGPT}.
\newblock \bibinfo{journal}{\emph{OpenAI Blog}} (\bibinfo{date}{11} \bibinfo{year}{2022}).
\newblock
\urldef\tempurl%
\url{https://openai.com/blog/chatgpt}
\showURL{%
\tempurl}
\newblock
\shownote{Retrieved 9/10/2023}.


\bibitem[Augenstein et~al\mbox{.}(2024)]%
        {augenstein2023factuality}
\bibfield{author}{\bibinfo{person}{Isabelle Augenstein}, \bibinfo{person}{Timothy Baldwin}, \bibinfo{person}{Meeyoung Cha}, \bibinfo{person}{Tanmoy Chakraborty}, \bibinfo{person}{Giovanni~Luca Ciampaglia}, \bibinfo{person}{David Corney}, \bibinfo{person}{Renee DiResta}, \bibinfo{person}{Emilio Ferrara}, \bibinfo{person}{Scott Hale}, \bibinfo{person}{Alon Halevy}, \bibinfo{person}{Eduard Hovy}, \bibinfo{person}{Heng Ji}, \bibinfo{person}{Filippo Menczer}, \bibinfo{person}{Ruben Miguez}, \bibinfo{person}{Preslav Nakov}, \bibinfo{person}{Dietram Scheufele}, \bibinfo{person}{Shivam Sharma}, {and} \bibinfo{person}{Giovanni Zagni}.} \bibinfo{year}{2024}\natexlab{}.
\newblock \showarticletitle{Factuality challenges in the era of large language models and opportunities for fact-checking}.
\newblock \bibinfo{journal}{\emph{Nature Machine Intelligence}} \bibinfo{volume}{6}, \bibinfo{number}{8} (\bibinfo{date}{Aug.} \bibinfo{year}{2024}), \bibinfo{pages}{852--863}.
\newblock
\showISSN{2522-5839}
\href{https://doi.org/10.1038/s42256-024-00881-z}{doi:\nolinkurl{10.1038/s42256-024-00881-z}}


\bibitem[Bernstein et~al\mbox{.}(2021)]%
        {bernstein14}
\bibfield{author}{\bibinfo{person}{Michael Bernstein}, \bibinfo{person}{Andrés Monroy-Hernández}, \bibinfo{person}{Drew Harry}, \bibinfo{person}{Paul André}, \bibinfo{person}{Katrina Panovich}, {and} \bibinfo{person}{Greg Vargas}.} \bibinfo{year}{2021}\natexlab{}.
\newblock \showarticletitle{4chan and /b/: {An} {Analysis} of {Anonymity} and {Ephemerality} in a {Large} {Online} {Community}}.
\newblock \bibinfo{journal}{\emph{Proceedings of the International AAAI Conference on Web and Social Media}} \bibinfo{volume}{5}, \bibinfo{number}{1} (\bibinfo{date}{Aug.} \bibinfo{year}{2021}), \bibinfo{pages}{50--57}.
\newblock
\href{https://doi.org/10.1609/icwsm.v5i1.14134}{doi:\nolinkurl{10.1609/icwsm.v5i1.14134}}
\newblock
\shownote{Section: Full Papers}.


\bibitem[Bozarth et~al\mbox{.}(2023)]%
        {Bozarth2023}
\bibfield{author}{\bibinfo{person}{Lia Bozarth}, \bibinfo{person}{Jane Im}, \bibinfo{person}{Christopher Quarles}, {and} \bibinfo{person}{Ceren Budak}.} \bibinfo{year}{2023}\natexlab{}.
\newblock \showarticletitle{Wisdom of Two Crowds: Misinformation Moderation on Reddit and How to Improve This Process---A Case Study of COVID-19}.
\newblock \bibinfo{journal}{\emph{Proc. ACM Hum.-Comput. Interact.}} \bibinfo{volume}{7}, \bibinfo{number}{CSCW1}, Article \bibinfo{articleno}{155} (\bibinfo{date}{4} \bibinfo{year}{2023}), \bibinfo{numpages}{33}~pages.
\newblock
\href{https://doi.org/10.1145/3579631}{doi:\nolinkurl{10.1145/3579631}}


\bibitem[Braun and Clarke(2019)]%
        {Braun19}
\bibfield{author}{\bibinfo{person}{Virginia Braun} {and} \bibinfo{person}{Victoria Clarke}.} \bibinfo{year}{2019}\natexlab{}.
\newblock \showarticletitle{Reflecting on reflexive thematic analysis}.
\newblock \bibinfo{journal}{\emph{Qualitative Research in Sport, Exercise and Health}} \bibinfo{volume}{11}, \bibinfo{number}{4} (\bibinfo{year}{2019}), \bibinfo{pages}{589--597}.
\newblock
\href{https://doi.org/10.1080/2159676X.2019.1628806}{doi:\nolinkurl{10.1080/2159676X.2019.1628806}}
\newblock
\shownote{Publisher: Routledge tex.eprint: https://doi.org/10.1080/2159676X.2019.1628806}.


\bibitem[Braun and Clarke(2021a)]%
        {braun21}
\bibfield{author}{\bibinfo{person}{Virginia Braun} {and} \bibinfo{person}{Victoria Clarke}.} \bibinfo{year}{2021}\natexlab{a}.
\newblock \showarticletitle{One size fits all? {What} counts as quality practice in (reflexive) thematic analysis?}
\newblock \bibinfo{journal}{\emph{Qualitative Research in Psychology}} \bibinfo{volume}{18}, \bibinfo{number}{3} (\bibinfo{date}{July} \bibinfo{year}{2021}), \bibinfo{pages}{328--352}.
\newblock
\showISSN{1478-0887}
\href{https://doi.org/10.1080/14780887.2020.1769238}{doi:\nolinkurl{10.1080/14780887.2020.1769238}}
\newblock
\shownote{Publisher: Routledge}.


\bibitem[Braun and Clarke(2021b)]%
        {braunSaturateNotSaturate2021}
\bibfield{author}{\bibinfo{person}{Virginia Braun} {and} \bibinfo{person}{Victoria Clarke}.} \bibinfo{year}{2021}\natexlab{b}.
\newblock \showarticletitle{To saturate or not to saturate? {Questioning} data saturation as a useful concept for thematic analysis and sample-size rationales}.
\newblock \bibinfo{journal}{\emph{Qualitative Research in Sport, Exercise and Health}} \bibinfo{volume}{13}, \bibinfo{number}{2} (\bibinfo{date}{March} \bibinfo{year}{2021}), \bibinfo{pages}{201--216}.
\newblock
\showISSN{2159-676X}
\href{https://doi.org/10.1080/2159676X.2019.1704846}{doi:\nolinkurl{10.1080/2159676X.2019.1704846}}
\newblock
\shownote{Publisher: Routledge \_eprint: https://doi.org/10.1080/2159676X.2019.1704846}.


\bibitem[Bruckman(2006)]%
        {Bruckman2006-BRUTST}
\bibfield{author}{\bibinfo{person}{Amy Bruckman}.} \bibinfo{year}{2006}\natexlab{}.
\newblock \showarticletitle{Teaching Students to Study Online Communities Ethically}.
\newblock \bibinfo{journal}{\emph{Journal of Information Ethics}} \bibinfo{volume}{15}, \bibinfo{number}{2} (\bibinfo{year}{2006}), \bibinfo{pages}{82--98}.
\newblock
\href{https://doi.org/10.3172/jie.15.2.82}{doi:\nolinkurl{10.3172/jie.15.2.82}}


\bibitem[Burtch et~al\mbox{.}(2024)]%
        {burtch_consequences_2024}
\bibfield{author}{\bibinfo{person}{Gordon Burtch}, \bibinfo{person}{Dokyun Lee}, {and} \bibinfo{person}{Zhichen Chen}.} \bibinfo{year}{2024}\natexlab{}.
\newblock \showarticletitle{The consequences of generative {AI} for online knowledge communities}.
\newblock \bibinfo{journal}{\emph{Scientific Reports}} \bibinfo{volume}{14}, \bibinfo{number}{1} (\bibinfo{date}{May} \bibinfo{year}{2024}), \bibinfo{pages}{10413}.
\newblock
\showISSN{2045-2322}
\href{https://doi.org/10.1038/s41598-024-61221-0}{doi:\nolinkurl{10.1038/s41598-024-61221-0}}
\newblock
\shownote{Publisher: Nature Publishing Group}.


\bibitem[Caine(2016)]%
        {caine16}
\bibfield{author}{\bibinfo{person}{Kelly Caine}.} \bibinfo{year}{2016}\natexlab{}.
\newblock \showarticletitle{Local Standards for Sample Size at CHI}. In \bibinfo{booktitle}{\emph{Proceedings of the 2016 CHI Conference on Human Factors in Computing Systems}} (San Jose, California, USA) \emph{(\bibinfo{series}{CHI '16})}. \bibinfo{publisher}{Association for Computing Machinery}, \bibinfo{address}{New York, NY, USA}, \bibinfo{pages}{981–992}.
\newblock
\showISBNx{9781450333627}
\href{https://doi.org/10.1145/2858036.2858498}{doi:\nolinkurl{10.1145/2858036.2858498}}


\bibitem[Chancellor et~al\mbox{.}(2018)]%
        {chancellor18}
\bibfield{author}{\bibinfo{person}{Stevie Chancellor}, \bibinfo{person}{Andrea Hu}, {and} \bibinfo{person}{Munmun De~Choudhury}.} \bibinfo{year}{2018}\natexlab{}.
\newblock \showarticletitle{Norms Matter: Contrasting Social Support Around Behavior Change in Online Weight Loss Communities}. In \bibinfo{booktitle}{\emph{Proceedings of the 2018 CHI Conference on Human Factors in Computing Systems}} (Montreal QC, Canada) \emph{(\bibinfo{series}{CHI '18})}. \bibinfo{publisher}{Association for Computing Machinery}, \bibinfo{address}{New York, NY, USA}, \bibinfo{pages}{1–14}.
\newblock
\showISBNx{9781450356206}
\href{https://doi.org/10.1145/3173574.3174240}{doi:\nolinkurl{10.1145/3173574.3174240}}


\bibitem[Chandrasekharan et~al\mbox{.}(2019)]%
        {Chandrasekharan19}
\bibfield{author}{\bibinfo{person}{Eshwar Chandrasekharan}, \bibinfo{person}{Chaitrali Gandhi}, \bibinfo{person}{Matthew~Wortley Mustelier}, {and} \bibinfo{person}{Eric Gilbert}.} \bibinfo{year}{2019}\natexlab{}.
\newblock \showarticletitle{Crossmod: A Cross-Community Learning-Based System to Assist Reddit Moderators}.
\newblock \bibinfo{journal}{\emph{Proc. ACM Hum.-Comput. Interact.}} \bibinfo{volume}{3}, \bibinfo{number}{CSCW}, Article \bibinfo{articleno}{174} (\bibinfo{date}{11} \bibinfo{year}{2019}), \bibinfo{numpages}{30}~pages.
\newblock
\href{https://doi.org/10.1145/3359276}{doi:\nolinkurl{10.1145/3359276}}


\bibitem[Chandrasekharan et~al\mbox{.}(2018)]%
        {Chandrasekharan18}
\bibfield{author}{\bibinfo{person}{Eshwar Chandrasekharan}, \bibinfo{person}{Mattia Samory}, \bibinfo{person}{Shagun Jhaver}, \bibinfo{person}{Hunter Charvat}, \bibinfo{person}{Amy Bruckman}, \bibinfo{person}{Cliff Lampe}, \bibinfo{person}{Jacob Eisenstein}, {and} \bibinfo{person}{Eric Gilbert}.} \bibinfo{year}{2018}\natexlab{}.
\newblock \showarticletitle{The Internet's Hidden Rules: An Empirical Study of Reddit Norm Violations at Micro, Meso, and Macro Scales}.
\newblock \bibinfo{journal}{\emph{Proc. ACM Hum.-Comput. Interact.}} \bibinfo{volume}{2}, \bibinfo{number}{CSCW}, Article \bibinfo{articleno}{32} (\bibinfo{date}{11} \bibinfo{year}{2018}), \bibinfo{numpages}{25}~pages.
\newblock
\href{https://doi.org/10.1145/3274301}{doi:\nolinkurl{10.1145/3274301}}


\bibitem[Cheng et~al\mbox{.}(2017)]%
        {Cheng17}
\bibfield{author}{\bibinfo{person}{Justin Cheng}, \bibinfo{person}{Michael Bernstein}, \bibinfo{person}{Cristian Danescu-Niculescu-Mizil}, {and} \bibinfo{person}{Jure Leskovec}.} \bibinfo{year}{2017}\natexlab{}.
\newblock \showarticletitle{Anyone Can Become a Troll: Causes of Trolling Behavior in Online Discussions}. In \bibinfo{booktitle}{\emph{Proceedings of the 2017 ACM Conference on Computer Supported Cooperative Work and Social Computing}} (Portland, Oregon, USA) \emph{(\bibinfo{series}{CSCW '17})}. \bibinfo{publisher}{Association for Computing Machinery}, \bibinfo{address}{New York, NY, USA}, \bibinfo{pages}{1217–1230}.
\newblock
\showISBNx{9781450343350}
\href{https://doi.org/10.1145/2998181.2998213}{doi:\nolinkurl{10.1145/2998181.2998213}}


\bibitem[Clark et~al\mbox{.}(2021)]%
        {clark21}
\bibfield{author}{\bibinfo{person}{Elizabeth Clark}, \bibinfo{person}{Tal August}, \bibinfo{person}{Sofia Serrano}, \bibinfo{person}{Nikita Haduong}, \bibinfo{person}{Suchin Gururangan}, {and} \bibinfo{person}{Noah~A. Smith}.} \bibinfo{year}{2021}\natexlab{}.
\newblock \showarticletitle{All {That}'s `{Human}' {Is} {Not} {Gold}: {Evaluating} {Human} {Evaluation} of {Generated} {Text}}. In \bibinfo{booktitle}{\emph{Proceedings of the 59th {Annual} {Meeting} of the {Association} for {Computational} {Linguistics} and the 11th {International} {Joint} {Conference} on {Natural} {Language} {Processing} ({Volume} 1: {Long} {Papers})}}. \bibinfo{publisher}{Association for Computational Linguistics}, \bibinfo{address}{Online}, \bibinfo{pages}{7282--7296}.
\newblock
\href{https://doi.org/10.18653/v1/2021.acl-long.565}{doi:\nolinkurl{10.18653/v1/2021.acl-long.565}}


\bibitem[Creswell and Creswell(2017)]%
        {creswellResearchDesignQualitative2017a}
\bibfield{author}{\bibinfo{person}{J.W. Creswell} {and} \bibinfo{person}{J.D. Creswell}.} \bibinfo{year}{2017}\natexlab{}.
\newblock \bibinfo{booktitle}{\emph{Research design: {Qualitative}, quantitative, and mixed methods approaches}}.
\newblock \bibinfo{publisher}{SAGE Publications}.
\newblock
\showISBNx{978-1-5063-8669-0}


\bibitem[Crothers et~al\mbox{.}(2023)]%
        {Crothers23}
\bibfield{author}{\bibinfo{person}{Evan~N. Crothers}, \bibinfo{person}{Nathalie Japkowicz}, {and} \bibinfo{person}{Herna~L. Viktor}.} \bibinfo{year}{2023}\natexlab{}.
\newblock \showarticletitle{Machine-Generated Text: A Comprehensive Survey of Threat Models and Detection Methods}.
\newblock \bibinfo{journal}{\emph{IEEE Access}}  \bibinfo{volume}{11} (\bibinfo{year}{2023}), \bibinfo{pages}{70977--71002}.
\newblock
\href{https://doi.org/10.1109/ACCESS.2023.3294090}{doi:\nolinkurl{10.1109/ACCESS.2023.3294090}}


\bibitem[Dimond et~al\mbox{.}(2012)]%
        {Dimond12}
\bibfield{author}{\bibinfo{person}{Jill~P. Dimond}, \bibinfo{person}{Casey Fiesler}, \bibinfo{person}{Betsy DiSalvo}, \bibinfo{person}{Jon Pelc}, {and} \bibinfo{person}{Amy~S. Bruckman}.} \bibinfo{year}{2012}\natexlab{}.
\newblock \showarticletitle{Qualitative data collection technologies: a comparison of instant messaging, email, and phone}. In \bibinfo{booktitle}{\emph{Proceedings of the 2012 ACM International Conference on Supporting Group Work}} (Sanibel Island, Florida, USA) \emph{(\bibinfo{series}{GROUP '12})}. \bibinfo{publisher}{Association for Computing Machinery}, \bibinfo{address}{New York, NY, USA}, \bibinfo{pages}{277–280}.
\newblock
\showISBNx{9781450314862}
\href{https://doi.org/10.1145/2389176.2389218}{doi:\nolinkurl{10.1145/2389176.2389218}}


\bibitem[DiResta and Goldstein(2024)]%
        {diresta2024spammers}
\bibfield{author}{\bibinfo{person}{Renée DiResta} {and} \bibinfo{person}{Josh~A. Goldstein}.} \bibinfo{year}{2024}\natexlab{}.
\newblock \showarticletitle{How spammers and scammers leverage {AI}-generated images on {Facebook} for audience growth}.
\newblock \bibinfo{journal}{\emph{Harvard Kennedy School Misinformation Review}} (\bibinfo{date}{Aug.} \bibinfo{year}{2024}).
\newblock
\href{https://doi.org/10.37016/mr-2020-151}{doi:\nolinkurl{10.37016/mr-2020-151}}


\bibitem[Dosono and Semaan(2019)]%
        {Dosono19}
\bibfield{author}{\bibinfo{person}{Bryan Dosono} {and} \bibinfo{person}{Bryan Semaan}.} \bibinfo{year}{2019}\natexlab{}.
\newblock \showarticletitle{Moderation Practices as Emotional Labor in Sustaining Online Communities: The Case of AAPI Identity Work on Reddit}. In \bibinfo{booktitle}{\emph{Proceedings of the 2019 CHI Conference on Human Factors in Computing Systems}} (Glasgow, Scotland Uk) \emph{(\bibinfo{series}{CHI '19})}. \bibinfo{publisher}{Association for Computing Machinery}, \bibinfo{address}{New York, NY, USA}, \bibinfo{pages}{1–13}.
\newblock
\showISBNx{9781450359702}
\href{https://doi.org/10.1145/3290605.3300372}{doi:\nolinkurl{10.1145/3290605.3300372}}


\bibitem[Edwards(2023)]%
        {edwards23}
\bibfield{author}{\bibinfo{person}{Benj Edwards}.} \bibinfo{year}{2023}\natexlab{}.
\newblock \bibinfo{title}{{OpenAI} discontinues its {AI} writing detector due to “low rate of accuracy”}.
\newblock
\urldef\tempurl%
\url{https://arstechnica.com/information-technology/2023/07/openai-discontinues-its-ai-writing-detector-due-to-low-rate-of-accuracy/}
\showURL{%
\tempurl}


\bibitem[Farid(2022)]%
        {farid_creating_2022}
\bibfield{author}{\bibinfo{person}{Hany Farid}.} \bibinfo{year}{2022}\natexlab{}.
\newblock \showarticletitle{Creating, {Using}, {Misusing}, and {Detecting} {Deep} {Fakes}}.
\newblock \bibinfo{journal}{\emph{Journal of Online Trust and Safety}} \bibinfo{volume}{1}, \bibinfo{number}{4} (\bibinfo{date}{Sept.} \bibinfo{year}{2022}).
\newblock
\showISSN{2770-3142}
\href{https://doi.org/10.54501/jots.v1i4.56}{doi:\nolinkurl{10.54501/jots.v1i4.56}}
\newblock
\shownote{Number: 4}.


\bibitem[Ferrara et~al\mbox{.}(2016)]%
        {Ferrara2016}
\bibfield{author}{\bibinfo{person}{Emilio Ferrara}, \bibinfo{person}{Onur Varol}, \bibinfo{person}{Clayton Davis}, \bibinfo{person}{Filippo Menczer}, {and} \bibinfo{person}{Alessandro Flammini}.} \bibinfo{year}{2016}\natexlab{}.
\newblock \showarticletitle{The Rise of Social Bots}.
\newblock \bibinfo{journal}{\emph{Commun. ACM}} \bibinfo{volume}{59}, \bibinfo{number}{7} (\bibinfo{date}{6} \bibinfo{year}{2016}), \bibinfo{pages}{96–104}.
\newblock
\showISSN{0001-0782}
\href{https://doi.org/10.1145/2818717}{doi:\nolinkurl{10.1145/2818717}}


\bibitem[Fiesler(2024)]%
        {fiesler_ai_2024}
\bibfield{author}{\bibinfo{person}{Casey Fiesler}.} \bibinfo{year}{2024}\natexlab{}.
\newblock \bibinfo{title}{{AI} chatbots are intruding into online communities where people are trying to connect with other humans}.
\newblock
\urldef\tempurl%
\url{http://theconversation.com/ai-chatbots-are-intruding-into-online-communities-where-people-are-trying-to-connect-with-other-humans-229473}
\showURL{%
\tempurl}


\bibitem[Fiesler et~al\mbox{.}(2018)]%
        {FieslerEtal2018rrc}
\bibfield{author}{\bibinfo{person}{Casey Fiesler}, \bibinfo{person}{Jialun Jiang}, \bibinfo{person}{Joshua McCann}, \bibinfo{person}{Kyle Frye}, {and} \bibinfo{person}{Jed Brubaker}.} \bibinfo{year}{2018}\natexlab{}.
\newblock \showarticletitle{Reddit {Rules}! {Characterizing} an {Ecosystem} of {Governance}}.
\newblock \bibinfo{journal}{\emph{Proceedings of the International AAAI Conference on Web and Social Media}} \bibinfo{volume}{12}, \bibinfo{number}{1} (\bibinfo{date}{June} \bibinfo{year}{2018}).
\newblock
\href{https://doi.org/10.1609/icwsm.v12i1.15033}{doi:\nolinkurl{10.1609/icwsm.v12i1.15033}}
\newblock
\shownote{Section: Full Papers}.


\bibitem[Fiesler et~al\mbox{.}(2024)]%
        {Fiesler24}
\bibfield{author}{\bibinfo{person}{Casey Fiesler}, \bibinfo{person}{Michael Zimmer}, \bibinfo{person}{Nicholas Proferes}, \bibinfo{person}{Sarah Gilbert}, {and} \bibinfo{person}{Naiyan Jones}.} \bibinfo{year}{2024}\natexlab{}.
\newblock \showarticletitle{Remember the Human: A Systematic Review of Ethical Considerations in Reddit Research}.
\newblock \bibinfo{journal}{\emph{Proc. ACM Hum.-Comput. Interact.}} \bibinfo{volume}{8}, \bibinfo{number}{GROUP}, Article \bibinfo{articleno}{5} (\bibinfo{date}{Feb.} \bibinfo{year}{2024}), \bibinfo{numpages}{33}~pages.
\newblock
\href{https://doi.org/10.1145/3633070}{doi:\nolinkurl{10.1145/3633070}}


\bibitem[Gertner(2023)]%
        {gertner_wikipedias_2023}
\bibfield{author}{\bibinfo{person}{Jon Gertner}.} \bibinfo{year}{2023}\natexlab{}.
\newblock \showarticletitle{Wikipedia’s {Moment} of {Truth}}.
\newblock \bibinfo{journal}{\emph{The New York Times}} (\bibinfo{date}{July} \bibinfo{year}{2023}).
\newblock
\showISSN{0362-4331}
\urldef\tempurl%
\url{https://www.nytimes.com/2023/07/18/magazine/wikipedia-ai-chatgpt.html}
\showURL{%
\tempurl}


\bibitem[Gilbert(2023)]%
        {gilbert23}
\bibfield{author}{\bibinfo{person}{Sarah Gilbert}.} \bibinfo{year}{2023}\natexlab{}.
\newblock \showarticletitle{Towards Intersectional Moderation: An Alternative Model of Moderation Built on Care and Power}.
\newblock \bibinfo{journal}{\emph{Proc. ACM Hum.-Comput. Interact.}} \bibinfo{volume}{7}, \bibinfo{number}{CSCW2}, Article \bibinfo{articleno}{256} (\bibinfo{date}{10} \bibinfo{year}{2023}), \bibinfo{numpages}{32}~pages.
\newblock
\href{https://doi.org/10.1145/3610047}{doi:\nolinkurl{10.1145/3610047}}


\bibitem[Gilbert(2020)]%
        {Gilbert20}
\bibfield{author}{\bibinfo{person}{Sarah~A. Gilbert}.} \bibinfo{year}{2020}\natexlab{}.
\newblock \showarticletitle{"I Run the World's Largest Historical Outreach Project and It's on a Cesspool of a Website." Moderating a Public Scholarship Site on Reddit: A Case Study of r/AskHistorians}.
\newblock \bibinfo{journal}{\emph{Proc. ACM Hum.-Comput. Interact.}} \bibinfo{volume}{4}, \bibinfo{number}{CSCW1}, Article \bibinfo{articleno}{19} (\bibinfo{date}{5} \bibinfo{year}{2020}), \bibinfo{numpages}{27}~pages.
\newblock
\href{https://doi.org/10.1145/3392822}{doi:\nolinkurl{10.1145/3392822}}


\bibitem[Goldstein et~al\mbox{.}(2023)]%
        {goldstein23}
\bibfield{author}{\bibinfo{person}{Josh~A. Goldstein}, \bibinfo{person}{Girish Sastry}, \bibinfo{person}{Micah Musser}, \bibinfo{person}{Renee DiResta}, \bibinfo{person}{Matthew Gentzel}, {and} \bibinfo{person}{Katerina Sedova}.} \bibinfo{year}{2023}\natexlab{}.
\newblock \bibinfo{title}{Generative {Language} {Models} and {Automated} {Influence} {Operations}: {Emerging} {Threats} and {Potential} {Mitigations}}.
\newblock
\href{https://doi.org/10.48550/arXiv.2301.04246}{doi:\nolinkurl{10.48550/arXiv.2301.04246}}
\newblock
\shownote{arXiv:2301.04246 [cs]}.


\bibitem[Grimmelmann(2015)]%
        {grimmelmann2015virtues}
\bibfield{author}{\bibinfo{person}{James Grimmelmann}.} \bibinfo{year}{2015}\natexlab{}.
\newblock \showarticletitle{The virtues of moderation}.
\newblock \bibinfo{journal}{\emph{Yale JL \& Tech.}}  \bibinfo{volume}{17} (\bibinfo{year}{2015}), \bibinfo{pages}{42}.
\newblock


\bibitem[Haimson et~al\mbox{.}(2021)]%
        {Haimson21}
\bibfield{author}{\bibinfo{person}{Oliver~L. Haimson}, \bibinfo{person}{Daniel Delmonaco}, \bibinfo{person}{Peipei Nie}, {and} \bibinfo{person}{Andrea Wegner}.} \bibinfo{year}{2021}\natexlab{}.
\newblock \showarticletitle{Disproportionate Removals and Differing Content Moderation Experiences for Conservative, Transgender, and Black Social Media Users: Marginalization and Moderation Gray Areas}.
\newblock \bibinfo{journal}{\emph{Proc. ACM Hum.-Comput. Interact.}} \bibinfo{volume}{5}, \bibinfo{number}{CSCW2}, Article \bibinfo{articleno}{466} (\bibinfo{date}{10} \bibinfo{year}{2021}), \bibinfo{numpages}{35}~pages.
\newblock
\href{https://doi.org/10.1145/3479610}{doi:\nolinkurl{10.1145/3479610}}


\bibitem[Han et~al\mbox{.}(2023)]%
        {han_hate_2023}
\bibfield{author}{\bibinfo{person}{Catherine Han}, \bibinfo{person}{Joseph Seering}, \bibinfo{person}{Deepak Kumar}, \bibinfo{person}{Jeffrey~T. Hancock}, {and} \bibinfo{person}{Zakir Durumeric}.} \bibinfo{year}{2023}\natexlab{}.
\newblock \showarticletitle{Hate {Raids} on {Twitch}: {Echoes} of the {Past}, {New} {Modalities}, and {Implications} for {Platform} {Governance}}.
\newblock \bibinfo{journal}{\emph{Proceedings of the ACM on Human-Computer Interaction}} \bibinfo{volume}{7}, \bibinfo{number}{CSCW1} (\bibinfo{date}{April} \bibinfo{year}{2023}), \bibinfo{pages}{133:1--133:28}.
\newblock
\href{https://doi.org/10.1145/3579609}{doi:\nolinkurl{10.1145/3579609}}


\bibitem[Hancock et~al\mbox{.}(2020)]%
        {hancock_aimc_2020}
\bibfield{author}{\bibinfo{person}{Jeffrey~T Hancock}, \bibinfo{person}{Mor Naaman}, {and} \bibinfo{person}{Karen Levy}.} \bibinfo{year}{2020}\natexlab{}.
\newblock \showarticletitle{{AI}-{Mediated} {Communication}: {Definition}, {Research} {Agenda}, and {Ethical} {Considerations}}.
\newblock \bibinfo{journal}{\emph{Journal of Computer-Mediated Communication}} \bibinfo{volume}{25}, \bibinfo{number}{1} (\bibinfo{date}{March} \bibinfo{year}{2020}), \bibinfo{pages}{89--100}.
\newblock
\showISSN{1083-6101}
\href{https://doi.org/10.1093/jcmc/zmz022}{doi:\nolinkurl{10.1093/jcmc/zmz022}}


\bibitem[Hanley and Durumeric(2024)]%
        {hanley_machinemade_2024}
\bibfield{author}{\bibinfo{person}{Hans W.~A. Hanley} {and} \bibinfo{person}{Zakir Durumeric}.} \bibinfo{year}{2024}\natexlab{}.
\newblock \showarticletitle{Machine-{Made} {Media}: {Monitoring} the {Mobilization} of {Machine}-{Generated} {Articles} on {Misinformation} and {Mainstream} {News} {Websites}}.
\newblock \bibinfo{journal}{\emph{Proceedings of the International AAAI Conference on Web and Social Media}}  \bibinfo{volume}{18} (\bibinfo{date}{May} \bibinfo{year}{2024}), \bibinfo{pages}{542--556}.
\newblock
\showISSN{2334-0770}
\href{https://doi.org/10.1609/icwsm.v18i1.31333}{doi:\nolinkurl{10.1609/icwsm.v18i1.31333}}


\bibitem[Im et~al\mbox{.}(2020)]%
        {Im20}
\bibfield{author}{\bibinfo{person}{Jane Im}, \bibinfo{person}{Sonali Tandon}, \bibinfo{person}{Eshwar Chandrasekharan}, \bibinfo{person}{Taylor Denby}, {and} \bibinfo{person}{Eric Gilbert}.} \bibinfo{year}{2020}\natexlab{}.
\newblock \showarticletitle{Synthesized Social Signals: Computationally-Derived Social Signals from Account Histories}. In \bibinfo{booktitle}{\emph{Proceedings of the 2020 CHI Conference on Human Factors in Computing Systems}} (Honolulu, HI, USA) \emph{(\bibinfo{series}{CHI '20})}. \bibinfo{publisher}{Association for Computing Machinery}, \bibinfo{address}{New York, NY, USA}, \bibinfo{pages}{1–12}.
\newblock
\showISBNx{9781450367080}
\href{https://doi.org/10.1145/3313831.3376383}{doi:\nolinkurl{10.1145/3313831.3376383}}


\bibitem[Jakesch et~al\mbox{.}(2019)]%
        {Jakesch19}
\bibfield{author}{\bibinfo{person}{Maurice Jakesch}, \bibinfo{person}{Megan French}, \bibinfo{person}{Xiao Ma}, \bibinfo{person}{Jeffrey~T. Hancock}, {and} \bibinfo{person}{Mor Naaman}.} \bibinfo{year}{2019}\natexlab{}.
\newblock \showarticletitle{AI-Mediated Communication: How the Perception That Profile Text Was Written by AI Affects Trustworthiness}. In \bibinfo{booktitle}{\emph{Proceedings of the 2019 CHI Conference on Human Factors in Computing Systems}} (Glasgow, Scotland Uk) \emph{(\bibinfo{series}{CHI '19})}. \bibinfo{publisher}{Association for Computing Machinery}, \bibinfo{address}{New York, NY, USA}, \bibinfo{pages}{1–13}.
\newblock
\showISBNx{9781450359702}
\href{https://doi.org/10.1145/3290605.3300469}{doi:\nolinkurl{10.1145/3290605.3300469}}


\bibitem[Jakesch et~al\mbox{.}(2023)]%
        {Jakesch_2023}
\bibfield{author}{\bibinfo{person}{Maurice Jakesch}, \bibinfo{person}{Jeffrey~T. Hancock}, {and} \bibinfo{person}{Mor Naaman}.} \bibinfo{year}{2023}\natexlab{}.
\newblock \showarticletitle{Human heuristics for {AI}-generated language are flawed}.
\newblock \bibinfo{journal}{\emph{Proceedings of the National Academy of Sciences}} \bibinfo{volume}{120}, \bibinfo{number}{11} (\bibinfo{date}{March} \bibinfo{year}{2023}), \bibinfo{pages}{e2208839120}.
\newblock
\href{https://doi.org/10.1073/pnas.2208839120}{doi:\nolinkurl{10.1073/pnas.2208839120}}
\newblock
\shownote{Publisher: Proceedings of the National Academy of Sciences}.


\bibitem[Jenkins(2006)]%
        {jenkins_introduction_2006}
\bibfield{author}{\bibinfo{person}{Henry Jenkins}.} \bibinfo{year}{2006}\natexlab{}.
\newblock \showarticletitle{Introduction: “{Worship} at the {Altar} of {Convergence}”: {A} {New} {Paradigm} for {Understanding} {Media} {Change}}.
\newblock In \bibinfo{booktitle}{\emph{Convergence {Culture}: {Where} {Old} and {New} {Media} {Collide}}}. \bibinfo{publisher}{NYU Press}, \bibinfo{pages}{20}.
\newblock
\showISBNx{978-0-8147-4281-5}
\urldef\tempurl%
\url{http://www.jstor.org/stable/j.ctt9qffwr.4}
\showURL{%
\tempurl}


\bibitem[Jhaver et~al\mbox{.}(2019a)]%
        {JhaverAppling19}
\bibfield{author}{\bibinfo{person}{Shagun Jhaver}, \bibinfo{person}{Darren~Scott Appling}, \bibinfo{person}{Eric Gilbert}, {and} \bibinfo{person}{Amy Bruckman}.} \bibinfo{year}{2019}\natexlab{a}.
\newblock \showarticletitle{"Did You Suspect the Post Would be Removed?": Understanding User Reactions to Content Removals on Reddit}.
\newblock \bibinfo{journal}{\emph{Proc. ACM Hum.-Comput. Interact.}} \bibinfo{volume}{3}, \bibinfo{number}{CSCW}, Article \bibinfo{articleno}{192} (\bibinfo{date}{11} \bibinfo{year}{2019}), \bibinfo{numpages}{33}~pages.
\newblock
\href{https://doi.org/10.1145/3359294}{doi:\nolinkurl{10.1145/3359294}}


\bibitem[Jhaver et~al\mbox{.}(2019b)]%
        {Jhaver19}
\bibfield{author}{\bibinfo{person}{Shagun Jhaver}, \bibinfo{person}{Iris Birman}, \bibinfo{person}{Eric Gilbert}, {and} \bibinfo{person}{Amy Bruckman}.} \bibinfo{year}{2019}\natexlab{b}.
\newblock \showarticletitle{Human-Machine Collaboration for Content Regulation: The Case of Reddit Automoderator}.
\newblock \bibinfo{journal}{\emph{ACM Trans. Comput.-Hum. Interact.}} \bibinfo{volume}{26}, \bibinfo{number}{5}, Article \bibinfo{articleno}{31} (\bibinfo{date}{7} \bibinfo{year}{2019}), \bibinfo{numpages}{35}~pages.
\newblock
\showISSN{1073-0516}
\href{https://doi.org/10.1145/3338243}{doi:\nolinkurl{10.1145/3338243}}


\bibitem[Kadoma et~al\mbox{.}(2025)]%
        {kadoma2024generativeaiperceptualharms}
\bibfield{author}{\bibinfo{person}{Kowe Kadoma}, \bibinfo{person}{Dana\'{e} Metaxa}, {and} \bibinfo{person}{Mor Naaman}.} \bibinfo{year}{2025}\natexlab{}.
\newblock \showarticletitle{Generative AI and Perceptual Harms: Who's Suspected of using LLMs?}. In \bibinfo{booktitle}{\emph{Proceedings of the 2025 CHI Conference on Human Factors in Computing Systems}} \emph{(\bibinfo{series}{CHI '25})}. \bibinfo{publisher}{Association for Computing Machinery}, \bibinfo{address}{New York, NY, USA}, Article \bibinfo{articleno}{861}, \bibinfo{numpages}{17}~pages.
\newblock
\showISBNx{9798400713941}
\href{https://doi.org/10.1145/3706598.3713897}{doi:\nolinkurl{10.1145/3706598.3713897}}


\bibitem[Kiene et~al\mbox{.}(2016)]%
        {Kiene16}
\bibfield{author}{\bibinfo{person}{Charles Kiene}, \bibinfo{person}{Andr\'{e}s Monroy-Hern\'{a}ndez}, {and} \bibinfo{person}{Benjamin~Mako Hill}.} \bibinfo{year}{2016}\natexlab{}.
\newblock \showarticletitle{Surviving an "Eternal September": How an Online Community Managed a Surge of Newcomers}. In \bibinfo{booktitle}{\emph{Proceedings of the 2016 CHI Conference on Human Factors in Computing Systems}} (San Jose, California, USA) \emph{(\bibinfo{series}{CHI '16})}. \bibinfo{publisher}{Association for Computing Machinery}, \bibinfo{address}{New York, NY, USA}, \bibinfo{pages}{1152–1156}.
\newblock
\showISBNx{9781450333627}
\href{https://doi.org/10.1145/2858036.2858356}{doi:\nolinkurl{10.1145/2858036.2858356}}


\bibitem[Kiesler et~al\mbox{.}(2011)]%
        {kiesler11}
\bibfield{author}{\bibinfo{person}{Sara Kiesler}, \bibinfo{person}{Robert Kraut}, \bibinfo{person}{Paul Resnick}, {and} \bibinfo{person}{Aniket Kittur}.} \bibinfo{year}{2011}\natexlab{}.
\newblock \showarticletitle{Regulating behavior in online communities}.
\newblock \bibinfo{journal}{\emph{Building successful online communities: Evidence-based social design}}  \bibinfo{volume}{1} (\bibinfo{year}{2011}), \bibinfo{pages}{125--179}.
\newblock


\bibitem[Kobak et~al\mbox{.}(2025)]%
        {kobakDelvingLLMassistedWriting2025}
\bibfield{author}{\bibinfo{person}{Dmitry Kobak}, \bibinfo{person}{Rita González-Márquez}, \bibinfo{person}{Emőke-Ágnes Horvát}, {and} \bibinfo{person}{Jan Lause}.} \bibinfo{year}{2025}\natexlab{}.
\newblock \showarticletitle{Delving into {LLM}-assisted writing in biomedical publications through excess vocabulary}.
\newblock \bibinfo{journal}{\emph{Science Advances}} \bibinfo{volume}{11}, \bibinfo{number}{27} (\bibinfo{date}{July} \bibinfo{year}{2025}), \bibinfo{pages}{eadt3813}.
\newblock
\href{https://doi.org/10.1126/sciadv.adt3813}{doi:\nolinkurl{10.1126/sciadv.adt3813}}
\newblock
\shownote{Publisher: American Association for the Advancement of Science}.


\bibitem[Koshy et~al\mbox{.}(2023)]%
        {Koshy23}
\bibfield{author}{\bibinfo{person}{Vinay Koshy}, \bibinfo{person}{Tanvi Bajpai}, \bibinfo{person}{Eshwar Chandrasekharan}, \bibinfo{person}{Hari Sundaram}, {and} \bibinfo{person}{Karrie Karahalios}.} \bibinfo{year}{2023}\natexlab{}.
\newblock \showarticletitle{Measuring User-Moderator Alignment on r/ChangeMyView}.
\newblock \bibinfo{journal}{\emph{Proc. ACM Hum.-Comput. Interact.}} \bibinfo{volume}{7}, \bibinfo{number}{CSCW2}, Article \bibinfo{articleno}{286} (\bibinfo{date}{10} \bibinfo{year}{2023}), \bibinfo{numpages}{36}~pages.
\newblock
\href{https://doi.org/10.1145/3610077}{doi:\nolinkurl{10.1145/3610077}}


\bibitem[Kraut and Resnick(2011)]%
        {kraut11}
\bibfield{author}{\bibinfo{person}{Robert~E Kraut} {and} \bibinfo{person}{Paul Resnick}.} \bibinfo{year}{2011}\natexlab{}.
\newblock \showarticletitle{Encouraging contribution to online communities}.
\newblock \bibinfo{journal}{\emph{Building successful online communities: Evidence-based social design}} (\bibinfo{year}{2011}), \bibinfo{pages}{21--76}.
\newblock


\bibitem[Lampe et~al\mbox{.}(2010)]%
        {Lampe10}
\bibfield{author}{\bibinfo{person}{Cliff Lampe}, \bibinfo{person}{Rick Wash}, \bibinfo{person}{Alcides Velasquez}, {and} \bibinfo{person}{Elif Ozkaya}.} \bibinfo{year}{2010}\natexlab{}.
\newblock \showarticletitle{Motivations to Participate in Online Communities}. In \bibinfo{booktitle}{\emph{Proceedings of the SIGCHI Conference on Human Factors in Computing Systems}} (Atlanta, Georgia, USA) \emph{(\bibinfo{series}{CHI '10})}. \bibinfo{publisher}{Association for Computing Machinery}, \bibinfo{address}{New York, NY, USA}, \bibinfo{pages}{1927–1936}.
\newblock
\showISBNx{9781605589299}
\href{https://doi.org/10.1145/1753326.1753616}{doi:\nolinkurl{10.1145/1753326.1753616}}


\bibitem[Lave and Wenger(1991)]%
        {lave1991situated}
\bibfield{author}{\bibinfo{person}{Jean Lave} {and} \bibinfo{person}{Etienne Wenger}.} \bibinfo{year}{1991}\natexlab{}.
\newblock \bibinfo{booktitle}{\emph{Situated learning: Legitimate peripheral participation}}.
\newblock \bibinfo{publisher}{Cambridge university press}.
\newblock


\bibitem[Lessig(1999)]%
        {Lessig99}
\bibfield{author}{\bibinfo{person}{Lawrence Lessig}.} \bibinfo{year}{1999}\natexlab{}.
\newblock \bibinfo{booktitle}{\emph{Code and Other Laws of Cyberspace}}.
\newblock \bibinfo{publisher}{Basic Books, Inc.}, \bibinfo{address}{USA}.
\newblock
\showISBNx{046503912X}


\bibitem[Li et~al\mbox{.}(2022)]%
        {li_all_2022}
\bibfield{author}{\bibinfo{person}{Hanlin Li}, \bibinfo{person}{Brent Hecht}, {and} \bibinfo{person}{Stevie Chancellor}.} \bibinfo{year}{2022}\natexlab{}.
\newblock \showarticletitle{All {That}’s {Happening} behind the {Scenes}: {Putting} the {Spotlight} on {Volunteer} {Moderator} {Labor} in {Reddit}}.
\newblock \bibinfo{journal}{\emph{Proceedings of the International AAAI Conference on Web and Social Media}}  \bibinfo{volume}{16} (\bibinfo{date}{May} \bibinfo{year}{2022}), \bibinfo{pages}{584--595}.
\newblock
\showISSN{2334-0770}
\href{https://doi.org/10.1609/icwsm.v16i1.19317}{doi:\nolinkurl{10.1609/icwsm.v16i1.19317}}


\bibitem[Liang et~al\mbox{.}(2024)]%
        {liang_monitoring_2024}
\bibfield{author}{\bibinfo{person}{Weixin Liang}, \bibinfo{person}{Zachary Izzo}, \bibinfo{person}{Yaohui Zhang}, \bibinfo{person}{Haley Lepp}, \bibinfo{person}{Hancheng Cao}, \bibinfo{person}{Xuandong Zhao}, \bibinfo{person}{Lingjiao Chen}, \bibinfo{person}{Haotian Ye}, \bibinfo{person}{Sheng Liu}, \bibinfo{person}{Zhi Huang}, \bibinfo{person}{Daniel~A. McFarland}, {and} \bibinfo{person}{James~Y. Zou}.} \bibinfo{year}{2024}\natexlab{}.
\newblock \showarticletitle{Monitoring AI-modified content at scale: a case study on the impact of ChatGPT on AI conference peer reviews}. In \bibinfo{booktitle}{\emph{Proceedings of the 41st International Conference on Machine Learning}} (Vienna, Austria) \emph{(\bibinfo{series}{ICML'24})}. \bibinfo{publisher}{JMLR.org}, Article \bibinfo{articleno}{1192}, \bibinfo{numpages}{46}~pages.
\newblock


\bibitem[Liang et~al\mbox{.}(2023)]%
        {liang23}
\bibfield{author}{\bibinfo{person}{Weixin Liang}, \bibinfo{person}{Mert Yuksekgonul}, \bibinfo{person}{Yining Mao}, \bibinfo{person}{Eric Wu}, {and} \bibinfo{person}{James Zou}.} \bibinfo{year}{2023}\natexlab{}.
\newblock \showarticletitle{{GPT} detectors are biased against non-native {English} writers}.
\newblock \bibinfo{journal}{\emph{Patterns}} \bibinfo{volume}{4}, \bibinfo{number}{7} (\bibinfo{date}{July} \bibinfo{year}{2023}).
\newblock
\showISSN{2666-3899}
\href{https://doi.org/10.1016/j.patter.2023.100779}{doi:\nolinkurl{10.1016/j.patter.2023.100779}}


\bibitem[Lloyd et~al\mbox{.}(2025)]%
        {lloydAIRulesCharacterizing2025}
\bibfield{author}{\bibinfo{person}{Travis Lloyd}, \bibinfo{person}{Jennah Gosciak}, \bibinfo{person}{Tung Nguyen}, {and} \bibinfo{person}{Mor Naaman}.} \bibinfo{year}{2025}\natexlab{}.
\newblock \showarticletitle{{AI} {Rules}? {Characterizing} {Reddit} {Community} {Policies} {Towards} {AI}-{Generated} {Content}}. In \bibinfo{booktitle}{\emph{Proceedings of the 2025 {CHI} {Conference} on {Human} {Factors} in {Computing} {Systems}}} \emph{(\bibinfo{series}{{CHI} '25})}. \bibinfo{publisher}{Association for Computing Machinery}, \bibinfo{address}{New York, NY, USA}, \bibinfo{pages}{1--19}.
\newblock
\showISBNx{979-8-4007-1394-1}
\href{https://doi.org/10.1145/3706598.3713292}{doi:\nolinkurl{10.1145/3706598.3713292}}


\bibitem[Lunden(2023)]%
        {lunden_linkedin_2023}
\bibfield{author}{\bibinfo{person}{Ingrid Lunden}.} \bibinfo{year}{2023}\natexlab{}.
\newblock \bibinfo{title}{{LinkedIn}, now at {1B} users, turns on {OpenAI}-powered reading and writing tools}.
\newblock
\urldef\tempurl%
\url{https://techcrunch.com/2023/11/01/linkedin-now-at-1b-users-turns-on-openai-powered-reading-and-writing-tools/}
\showURL{%
\tempurl}


\bibitem[Makyen(2023)]%
        {makyen23}
\bibfield{author}{\bibinfo{person}{Makyen}.} \bibinfo{year}{2023}\natexlab{}.
\newblock \bibinfo{title}{Temporary policy: {Generative} {AI} (e.g., {ChatGPT}) is banned}.
\newblock
\urldef\tempurl%
\url{https://meta.stackoverflow.com/q/421831/9737437}
\showURL{%
\tempurl}


\bibitem[Matatov et~al\mbox{.}(2025)]%
        {matatovExaminingPrevalenceDynamics2025}
\bibfield{author}{\bibinfo{person}{Hana Matatov}, \bibinfo{person}{Marianne Aubin~Le Quéré}, \bibinfo{person}{Ofra Amir}, {and} \bibinfo{person}{Mor Naaman}.} \bibinfo{year}{2025}\natexlab{}.
\newblock \bibinfo{title}{Examining the {Prevalence} and {Dynamics} of {AI}-{Generated} {Media} in {Art} {Subreddits}}.
\newblock
\href{https://doi.org/10.48550/arXiv.2410.07302}{doi:\nolinkurl{10.48550/arXiv.2410.07302}}


\bibitem[Matias(2019a)]%
        {Matias19}
\bibfield{author}{\bibinfo{person}{J.~Nathan Matias}.} \bibinfo{year}{2019}\natexlab{a}.
\newblock \showarticletitle{The Civic Labor of Volunteer Moderators Online}.
\newblock \bibinfo{journal}{\emph{Social Media + Society}} \bibinfo{volume}{5}, \bibinfo{number}{2} (\bibinfo{year}{2019}), \bibinfo{pages}{2056305119836778}.
\newblock
\showeprint{https://doi.org/10.1177/2056305119836778}
\href{https://doi.org/10.1177/2056305119836778}{doi:\nolinkurl{10.1177/2056305119836778}}


\bibitem[Matias(2019b)]%
        {matias2019preventing}
\bibfield{author}{\bibinfo{person}{J~Nathan Matias}.} \bibinfo{year}{2019}\natexlab{b}.
\newblock \showarticletitle{Preventing harassment and increasing group participation through social norms in 2,190 online science discussions}.
\newblock \bibinfo{journal}{\emph{Proceedings of the National Academy of Sciences}} \bibinfo{volume}{116}, \bibinfo{number}{20} (\bibinfo{year}{2019}), \bibinfo{pages}{9785--9789}.
\newblock


\bibitem[Matias and Mou(2018)]%
        {Matias18}
\bibfield{author}{\bibinfo{person}{J.~Nathan Matias} {and} \bibinfo{person}{Merry Mou}.} \bibinfo{year}{2018}\natexlab{}.
\newblock \showarticletitle{CivilServant: Community-Led Experiments in Platform Governance}. In \bibinfo{booktitle}{\emph{Proceedings of the 2018 CHI Conference on Human Factors in Computing Systems}} (Montreal QC, Canada) \emph{(\bibinfo{series}{CHI '18})}. \bibinfo{publisher}{Association for Computing Machinery}, \bibinfo{address}{New York, NY, USA}, \bibinfo{pages}{1–13}.
\newblock
\showISBNx{9781450356206}
\href{https://doi.org/10.1145/3173574.3173583}{doi:\nolinkurl{10.1145/3173574.3173583}}


\bibitem[McDonald et~al\mbox{.}(2019)]%
        {mcdonald19}
\bibfield{author}{\bibinfo{person}{Nora McDonald}, \bibinfo{person}{Sarita Schoenebeck}, {and} \bibinfo{person}{Andrea Forte}.} \bibinfo{year}{2019}\natexlab{}.
\newblock \showarticletitle{Reliability and Inter-Rater Reliability in Qualitative Research: Norms and Guidelines for CSCW and HCI Practice}.
\newblock \bibinfo{journal}{\emph{Proc. ACM Hum.-Comput. Interact.}} \bibinfo{volume}{3}, \bibinfo{number}{CSCW}, Article \bibinfo{articleno}{72} (\bibinfo{date}{11} \bibinfo{year}{2019}), \bibinfo{numpages}{23}~pages.
\newblock
\href{https://doi.org/10.1145/3359174}{doi:\nolinkurl{10.1145/3359174}}


\bibitem[Menczer et~al\mbox{.}(2023)]%
        {menczer_addressing_2023}
\bibfield{author}{\bibinfo{person}{Filippo Menczer}, \bibinfo{person}{David Crandall}, \bibinfo{person}{Yong-Yeol Ahn}, {and} \bibinfo{person}{Apu Kapadia}.} \bibinfo{year}{2023}\natexlab{}.
\newblock \showarticletitle{Addressing the harms of {AI}-generated inauthentic content}.
\newblock \bibinfo{journal}{\emph{Nature Machine Intelligence}} \bibinfo{volume}{5}, \bibinfo{number}{7} (\bibinfo{date}{July} \bibinfo{year}{2023}), \bibinfo{pages}{679--680}.
\newblock
\showISSN{2522-5839}
\href{https://doi.org/10.1038/s42256-023-00690-w}{doi:\nolinkurl{10.1038/s42256-023-00690-w}}
\newblock
\shownote{Number: 7 Publisher: Nature Publishing Group}.


\bibitem[Mink et~al\mbox{.}(2024)]%
        {mink24}
\bibfield{author}{\bibinfo{person}{Jaron Mink}, \bibinfo{person}{Miranda Wei}, \bibinfo{person}{Collins~W. Munyendo}, \bibinfo{person}{Kurt Hugenberg}, \bibinfo{person}{Tadayoshi Kohno}, \bibinfo{person}{Elissa~M. Redmiles}, {and} \bibinfo{person}{Gang Wang}.} \bibinfo{year}{2024}\natexlab{}.
\newblock \showarticletitle{It's Trying Too Hard To Look Real: Deepfake Moderation Mistakes and Identity-Based Bias}. In \bibinfo{booktitle}{\emph{Proceedings of the CHI Conference on Human Factors in Computing Systems}} \emph{(\bibinfo{series}{CHI '24})}. \bibinfo{publisher}{Association for Computing Machinery}, \bibinfo{address}{New York, NY, USA}, Article \bibinfo{articleno}{778}, \bibinfo{numpages}{20}~pages.
\newblock
\href{https://doi.org/10.1145/3613904.3641999}{doi:\nolinkurl{10.1145/3613904.3641999}}


\bibitem[Nightingale and Farid(2022)]%
        {nightingale22}
\bibfield{author}{\bibinfo{person}{Sophie~J. Nightingale} {and} \bibinfo{person}{Hany Farid}.} \bibinfo{year}{2022}\natexlab{}.
\newblock \showarticletitle{{AI}-synthesized faces are indistinguishable from real faces and more trustworthy}.
\newblock \bibinfo{journal}{\emph{Proceedings of the National Academy of Sciences}} \bibinfo{volume}{119}, \bibinfo{number}{8} (\bibinfo{date}{Feb.} \bibinfo{year}{2022}), \bibinfo{pages}{e2120481119}.
\newblock
\href{https://doi.org/10.1073/pnas.2120481119}{doi:\nolinkurl{10.1073/pnas.2120481119}}
\newblock
\shownote{Publisher: Proceedings of the National Academy of Sciences}.


\bibitem[Oak and Shafiq(2024)]%
        {Oak24}
\bibfield{author}{\bibinfo{person}{Rajvardhan Oak} {and} \bibinfo{person}{Zubair Shafiq}.} \bibinfo{year}{2024}\natexlab{}.
\newblock \showarticletitle{Understanding Underground Incentivized Review Services}. In \bibinfo{booktitle}{\emph{Proceedings of the CHI Conference on Human Factors in Computing Systems}} \emph{(\bibinfo{series}{CHI '24})}. \bibinfo{publisher}{Association for Computing Machinery}, \bibinfo{address}{New York, NY, USA}, Article \bibinfo{articleno}{950}, \bibinfo{numpages}{18}~pages.
\newblock
\href{https://doi.org/10.1145/3613904.3642342}{doi:\nolinkurl{10.1145/3613904.3642342}}


\bibitem[Obar and Oeldorf-Hirsch(2020)]%
        {obar_biggest_2020}
\bibfield{author}{\bibinfo{person}{Jonathan~A. Obar} {and} \bibinfo{person}{Anne Oeldorf-Hirsch}.} \bibinfo{year}{2020}\natexlab{}.
\newblock \showarticletitle{The biggest lie on the {Internet}: ignoring the privacy policies and terms of service policies of social networking services}.
\newblock \bibinfo{journal}{\emph{Information, Communication \& Society}} \bibinfo{volume}{23}, \bibinfo{number}{1} (\bibinfo{date}{Jan.} \bibinfo{year}{2020}), \bibinfo{pages}{128--147}.
\newblock
\showISSN{1369-118X}
\href{https://doi.org/10.1080/1369118X.2018.1486870}{doi:\nolinkurl{10.1080/1369118X.2018.1486870}}


\bibitem[Park et~al\mbox{.}(2024)]%
        {park_ai_2024}
\bibfield{author}{\bibinfo{person}{Peter~S. Park}, \bibinfo{person}{Simon Goldstein}, \bibinfo{person}{Aidan O’Gara}, \bibinfo{person}{Michael Chen}, {and} \bibinfo{person}{Dan Hendrycks}.} \bibinfo{year}{2024}\natexlab{}.
\newblock \showarticletitle{{AI} deception: {A} survey of examples, risks, and potential solutions}.
\newblock \bibinfo{journal}{\emph{Patterns}} \bibinfo{volume}{5}, \bibinfo{number}{5} (\bibinfo{date}{May} \bibinfo{year}{2024}), \bibinfo{pages}{100988}.
\newblock
\showISSN{26663899}
\href{https://doi.org/10.1016/j.patter.2024.100988}{doi:\nolinkurl{10.1016/j.patter.2024.100988}}


\bibitem[Preece(1999)]%
        {preece_empathic_1999}
\bibfield{author}{\bibinfo{person}{Jenny Preece}.} \bibinfo{year}{1999}\natexlab{}.
\newblock \showarticletitle{Empathic communities: balancing emotional and factual communication}.
\newblock \bibinfo{journal}{\emph{Interacting with Computers}} \bibinfo{volume}{12}, \bibinfo{number}{1} (\bibinfo{date}{Sept.} \bibinfo{year}{1999}), \bibinfo{pages}{63--77}.
\newblock
\showISSN{0953-5438}
\href{https://doi.org/10.1016/S0953-5438(98)00056-3}{doi:\nolinkurl{10.1016/S0953-5438(98)00056-3}}


\bibitem[Prinster et~al\mbox{.}(2024)]%
        {Prinster24}
\bibfield{author}{\bibinfo{person}{Gale~H. Prinster}, \bibinfo{person}{C.~Estelle Smith}, \bibinfo{person}{Chenhao Tan}, {and} \bibinfo{person}{Brian~C. Keegan}.} \bibinfo{year}{2024}\natexlab{}.
\newblock \showarticletitle{Community Archetypes: An Empirical Framework for Guiding Research Methodologies to Reflect User Experiences of Sense of Virtual Community on Reddit}.
\newblock \bibinfo{journal}{\emph{Proc. ACM Hum.-Comput. Interact.}} \bibinfo{volume}{8}, \bibinfo{number}{CSCW1}, Article \bibinfo{articleno}{33} (\bibinfo{date}{April} \bibinfo{year}{2024}), \bibinfo{numpages}{33}~pages.
\newblock
\href{https://doi.org/10.1145/3637310}{doi:\nolinkurl{10.1145/3637310}}


\bibitem[Proferes et~al\mbox{.}(2021)]%
        {proferes21}
\bibfield{author}{\bibinfo{person}{Nicholas Proferes}, \bibinfo{person}{Naiyan Jones}, \bibinfo{person}{Sarah Gilbert}, \bibinfo{person}{Casey Fiesler}, {and} \bibinfo{person}{Michael Zimmer}.} \bibinfo{year}{2021}\natexlab{}.
\newblock \showarticletitle{Studying reddit: a systematic overview of disciplines, approaches, methods, and ethics}.
\newblock \bibinfo{journal}{\emph{Social Media + Society}} \bibinfo{volume}{7}, \bibinfo{number}{2} (\bibinfo{year}{2021}), \bibinfo{pages}{20563051211019004}.
\newblock
\href{https://doi.org/10.1177/20563051211019004}{doi:\nolinkurl{10.1177/20563051211019004}}
\newblock
\shownote{tex.eprint: https://doi.org/10.1177/20563051211019004}.


\bibitem[Rae(2024)]%
        {rae_effects_2024}
\bibfield{author}{\bibinfo{person}{Irene Rae}.} \bibinfo{year}{2024}\natexlab{}.
\newblock \showarticletitle{The {Effects} of {Perceived} {AI} {Use} {On} {Content} {Perceptions}}. In \bibinfo{booktitle}{\emph{Proceedings of the {CHI} {Conference} on {Human} {Factors} in {Computing} {Systems}}} \emph{(\bibinfo{series}{{CHI} '24})}. \bibinfo{publisher}{Association for Computing Machinery}, \bibinfo{address}{New York, NY, USA}, \bibinfo{pages}{1--14}.
\newblock
\href{https://doi.org/10.1145/3613904.3642076}{doi:\nolinkurl{10.1145/3613904.3642076}}


\bibitem[Ramesh et~al\mbox{.}(2021)]%
        {ramesh2021zeroshot}
\bibfield{author}{\bibinfo{person}{Aditya Ramesh}, \bibinfo{person}{Mikhail Pavlov}, \bibinfo{person}{Gabriel Goh}, \bibinfo{person}{Scott Gray}, \bibinfo{person}{Chelsea Voss}, \bibinfo{person}{Alec Radford}, \bibinfo{person}{Mark Chen}, {and} \bibinfo{person}{Ilya Sutskever}.} \bibinfo{year}{2021}\natexlab{}.
\newblock \bibinfo{title}{Zero-Shot Text-to-Image Generation}.
\newblock
\showeprint[arxiv]{2102.12092}~[cs.CV]


\bibitem[r/changemyview(2025)]%
        {changemyviewMETAUnauthorizedExperiment2025}
\bibfield{author}{\bibinfo{person}{Mod~Team r/changemyview}.} \bibinfo{year}{2025}\natexlab{}.
\newblock \bibinfo{title}{{META}: {Unauthorized} {Experiment} on {CMV} {Involving} {AI}-generated {Comments}}.
\newblock
\urldef\tempurl%
\url{https://www.reddit.com/r/changemyview/comments/1k8b2hj/meta_unauthorized_experiment_on_cmv_involving/}
\showURL{%
\tempurl}


\bibitem[Ren et~al\mbox{.}(2012)]%
        {ren_encouraging_2012}
\bibfield{author}{\bibinfo{person}{Yuqing Ren}, \bibinfo{person}{Robert Kraut}, \bibinfo{person}{Sara Kiesler}, {and} \bibinfo{person}{Paul Resnick}.} \bibinfo{year}{2012}\natexlab{}.
\newblock \showarticletitle{Encouraging commitment in online communities}.
\newblock \bibinfo{journal}{\emph{Building successful online communities: Evidence-based social design}} (\bibinfo{year}{2012}), \bibinfo{pages}{77--124}.
\newblock
\newblock
\shownote{Publisher: MIT Press Cambridge}.


\bibitem[Resnick and Zeckhauser(2002)]%
        {resnick2002trust}
\bibfield{author}{\bibinfo{person}{Paul Resnick} {and} \bibinfo{person}{Richard Zeckhauser}.} \bibinfo{year}{2002}\natexlab{}.
\newblock \showarticletitle{Trust among strangers in Internet transactions: Empirical analysis of eBay's reputation system}.
\newblock In \bibinfo{booktitle}{\emph{The Economics of the Internet and E-commerce}}. \bibinfo{publisher}{Emerald Group Publishing Limited}, \bibinfo{pages}{127--157}.
\newblock


\bibitem[Rini and Cohen(2022)]%
        {rini_deepfakes_2022}
\bibfield{author}{\bibinfo{person}{Regina Rini} {and} \bibinfo{person}{Leah Cohen}.} \bibinfo{year}{2022}\natexlab{}.
\newblock \showarticletitle{Deepfakes, {Deep} {Harms}}.
\newblock \bibinfo{journal}{\emph{Journal of Ethics and Social Philosophy}} \bibinfo{volume}{22}, \bibinfo{number}{2} (\bibinfo{year}{2022}), \bibinfo{pages}{143--161}.
\newblock
\urldef\tempurl%
\url{https://heinonline.org/HOL/P?h=hein.journals/jetshy22&i=150}
\showURL{%
\tempurl}


\bibitem[Sadasivan et~al\mbox{.}(2025)]%
        {sadasivan2023}
\bibfield{author}{\bibinfo{person}{Vinu~Sankar Sadasivan}, \bibinfo{person}{Aounon Kumar}, \bibinfo{person}{Sriram Balasubramanian}, \bibinfo{person}{Wenxiao Wang}, {and} \bibinfo{person}{Soheil Feizi}.} \bibinfo{year}{2025}\natexlab{}.
\newblock \showarticletitle{Can {AI}-Generated Text be Reliably Detected? Stress Testing {AI} Text Detectors Under Various Attacks}.
\newblock \bibinfo{journal}{\emph{Transactions on Machine Learning Research}} (\bibinfo{year}{2025}).
\newblock
\showISSN{2835-8856}
\urldef\tempurl%
\url{https://openreview.net/forum?id=OOgsAZdFOt}
\showURL{%
\tempurl}


\bibitem[Sanatizadeh et~al\mbox{.}(2025)]%
        {sanatizadeh_exploring_2023}
\bibfield{author}{\bibinfo{person}{Aida Sanatizadeh}, \bibinfo{person}{Yingda Lu}, \bibinfo{person}{Keran Zhao}, {and} \bibinfo{person}{Yuheng Hu}.} \bibinfo{year}{2025}\natexlab{}.
\newblock \showarticletitle{Engagement or entanglement? {The} dual impact of generative artificial intelligence in online knowledge exchange platforms}.
\newblock \bibinfo{journal}{\emph{Information \& Management}} \bibinfo{volume}{62}, \bibinfo{number}{6} (\bibinfo{date}{Sept.} \bibinfo{year}{2025}), \bibinfo{pages}{104178}.
\newblock
\showISSN{0378-7206}
\href{https://doi.org/10.1016/j.im.2025.104178}{doi:\nolinkurl{10.1016/j.im.2025.104178}}


\bibitem[Schluger et~al\mbox{.}(2022)]%
        {Schluger22}
\bibfield{author}{\bibinfo{person}{Charlotte Schluger}, \bibinfo{person}{Jonathan~P. Chang}, \bibinfo{person}{Cristian Danescu-Niculescu-Mizil}, {and} \bibinfo{person}{Karen Levy}.} \bibinfo{year}{2022}\natexlab{}.
\newblock \showarticletitle{Proactive Moderation of Online Discussions: Existing Practices and the Potential for Algorithmic Support}.
\newblock \bibinfo{journal}{\emph{Proc. ACM Hum.-Comput. Interact.}} \bibinfo{volume}{6}, \bibinfo{number}{CSCW2}, Article \bibinfo{articleno}{370} (\bibinfo{date}{11} \bibinfo{year}{2022}), \bibinfo{numpages}{27}~pages.
\newblock
\href{https://doi.org/10.1145/3555095}{doi:\nolinkurl{10.1145/3555095}}


\bibitem[Seering(2020)]%
        {Seering20}
\bibfield{author}{\bibinfo{person}{Joseph Seering}.} \bibinfo{year}{2020}\natexlab{}.
\newblock \showarticletitle{Reconsidering Self-Moderation: The Role of Research in Supporting Community-Based Models for Online Content Moderation}.
\newblock \bibinfo{journal}{\emph{Proc. ACM Hum.-Comput. Interact.}} \bibinfo{volume}{4}, \bibinfo{number}{CSCW2}, Article \bibinfo{articleno}{107} (\bibinfo{date}{11} \bibinfo{year}{2020}), \bibinfo{numpages}{28}~pages.
\newblock
\href{https://doi.org/10.1145/3415178}{doi:\nolinkurl{10.1145/3415178}}


\bibitem[Seering et~al\mbox{.}(2019)]%
        {SeeringEtal2019mec}
\bibfield{author}{\bibinfo{person}{Joseph Seering}, \bibinfo{person}{Tony Wang}, \bibinfo{person}{Jina Yoon}, {and} \bibinfo{person}{Geoff Kaufman}.} \bibinfo{year}{2019}\natexlab{}.
\newblock \showarticletitle{Moderator engagement and community development in the age of algorithms}.
\newblock \bibinfo{journal}{\emph{New Media \& Society}} \bibinfo{volume}{21}, \bibinfo{number}{7} (\bibinfo{date}{July} \bibinfo{year}{2019}), \bibinfo{pages}{1417--1443}.
\newblock
\showISSN{1461-4448}
\href{https://doi.org/10.1177/1461444818821316}{doi:\nolinkurl{10.1177/1461444818821316}}
\newblock
\shownote{Publisher: SAGE Publications}.


\bibitem[Smith et~al\mbox{.}(2022)]%
        {Smith22}
\bibfield{author}{\bibinfo{person}{C.~Estelle Smith}, \bibinfo{person}{Irfanul Alam}, \bibinfo{person}{Chenhao Tan}, \bibinfo{person}{Brian~C. Keegan}, {and} \bibinfo{person}{Anita~L. Blanchard}.} \bibinfo{year}{2022}\natexlab{}.
\newblock \showarticletitle{The Impact of Governance Bots on Sense of Virtual Community: Development and Validation of the GOV-BOTs Scale}.
\newblock \bibinfo{journal}{\emph{Proc. ACM Hum.-Comput. Interact.}} \bibinfo{volume}{6}, \bibinfo{number}{CSCW2}, Article \bibinfo{articleno}{462} (\bibinfo{date}{Nov.} \bibinfo{year}{2022}), \bibinfo{numpages}{30}~pages.
\newblock
\href{https://doi.org/10.1145/3555563}{doi:\nolinkurl{10.1145/3555563}}


\bibitem[Smith et~al\mbox{.}(2023)]%
        {Smith23}
\bibfield{author}{\bibinfo{person}{C.~Estelle Smith}, \bibinfo{person}{Hannah Miller~Hillberg}, {and} \bibinfo{person}{Zachary Levonian}.} \bibinfo{year}{2023}\natexlab{}.
\newblock \showarticletitle{"Thoughts \& Prayers" or ":Heart Reaction: \& :Prayer Reaction:": How the Release of New Reactions on CaringBridge Reshapes Supportive Communication in Health Crises}.
\newblock \bibinfo{journal}{\emph{Proc. ACM Hum.-Comput. Interact.}} \bibinfo{volume}{7}, \bibinfo{number}{CSCW2}, Article \bibinfo{articleno}{244} (\bibinfo{date}{Oct.} \bibinfo{year}{2023}), \bibinfo{numpages}{39}~pages.
\newblock
\href{https://doi.org/10.1145/3610035}{doi:\nolinkurl{10.1145/3610035}}


\bibitem[Thach et~al\mbox{.}(2024)]%
        {Thach24}
\bibfield{author}{\bibinfo{person}{Hibby Thach}, \bibinfo{person}{Samuel Mayworm}, \bibinfo{person}{Daniel Delmonaco}, {and} \bibinfo{person}{Oliver Haimson}.} \bibinfo{year}{2024}\natexlab{}.
\newblock \showarticletitle{(In)visible moderation: A digital ethnography of marginalized users and content moderation on Twitch and Reddit}.
\newblock \bibinfo{journal}{\emph{New Media \& Society}} \bibinfo{volume}{26}, \bibinfo{number}{7} (\bibinfo{year}{2024}), \bibinfo{pages}{4034--4055}.
\newblock
\href{https://doi.org/10.1177/14614448221109804}{doi:\nolinkurl{10.1177/14614448221109804}}


\bibitem[Thompson(2023)]%
        {Thompson23Discovered}
\bibfield{author}{\bibinfo{person}{Stuart~A. Thompson}.} \bibinfo{year}{2023}\natexlab{}.
\newblock \showarticletitle{{A.I}.-Generated Content Discovered on News Sites, Content Farms and Product Reviews}.
\newblock \bibinfo{journal}{\emph{The New York Times}} (\bibinfo{date}{5} \bibinfo{year}{2023}).
\newblock
\urldef\tempurl%
\url{https://www.nytimes.com/2023/05/19/technology/ai-generated-content-discovered-on-news-sites-content-farms-and-product-reviews.html}
\showURL{%
\tempurl}
\newblock
\shownote{Retrieved 9/10/2023}.


\bibitem[u/spez(2025)]%
        {spezRedditsNextChapter2025}
\bibfield{author}{\bibinfo{person}{u/spez}.} \bibinfo{year}{2025}\natexlab{}.
\newblock \bibinfo{title}{Reddit’s next chapter: smarter, easier, still human}.
\newblock
\urldef\tempurl%
\url{https://www.reddit.com/user/spez/comments/1kfciml/reddits_next_chapter_smarter_easier_still_human/}
\showURL{%
\tempurl}


\bibitem[Veselovsky et~al\mbox{.}(2025)]%
        {veselovsky_prevalence_2023}
\bibfield{author}{\bibinfo{person}{Veniamin Veselovsky}, \bibinfo{person}{Manoel Horta~Ribeiro}, \bibinfo{person}{Philip~J. Cozzolino}, \bibinfo{person}{Andrew Gordon}, \bibinfo{person}{David Rothschild}, {and} \bibinfo{person}{Robert West}.} \bibinfo{year}{2025}\natexlab{}.
\newblock \showarticletitle{Prevalence and Prevention of Large Language Model Use in Crowd Work}.
\newblock \bibinfo{journal}{\emph{Commun. ACM}} \bibinfo{volume}{68}, \bibinfo{number}{3} (\bibinfo{date}{Feb.} \bibinfo{year}{2025}), \bibinfo{pages}{42–47}.
\newblock
\showISSN{0001-0782}
\href{https://doi.org/10.1145/3685527}{doi:\nolinkurl{10.1145/3685527}}


\bibitem[Wang and Zheng(2024)]%
        {wang_can_2024}
\bibfield{author}{\bibinfo{person}{Xiaoxiao~(Shawn) Wang} {and} \bibinfo{person}{Jinyang Zheng}.} \bibinfo{year}{2024}\natexlab{}.
\newblock \bibinfo{title}{Can {Banning} {AI}-generated {Content} {Save} {User}-{Generated} {Q}\&{A} {Platforms}?}
\newblock
\href{https://doi.org/10.2139/ssrn.4750326}{doi:\nolinkurl{10.2139/ssrn.4750326}}


\bibitem[Weber-Wulff et~al\mbox{.}(2023)]%
        {weberwulff23}
\bibfield{author}{\bibinfo{person}{Debora Weber-Wulff}, \bibinfo{person}{Alla Anohina-Naumeca}, \bibinfo{person}{Sonja Bjelobaba}, \bibinfo{person}{Tomáš Foltýnek}, \bibinfo{person}{Jean Guerrero-Dib}, \bibinfo{person}{Olumide Popoola}, \bibinfo{person}{Petr Šigut}, {and} \bibinfo{person}{Lorna Waddington}.} \bibinfo{year}{2023}\natexlab{}.
\newblock \showarticletitle{Testing of detection tools for {AI}-generated text}.
\newblock \bibinfo{journal}{\emph{International Journal for Educational Integrity}} \bibinfo{volume}{19}, \bibinfo{number}{1} (\bibinfo{date}{Dec.} \bibinfo{year}{2023}), \bibinfo{pages}{26}.
\newblock
\showISSN{1833-2595}
\href{https://doi.org/10.1007/s40979-023-00146-z}{doi:\nolinkurl{10.1007/s40979-023-00146-z}}


\bibitem[Wei and Tyson(2024)]%
        {wei2024understandingimpactaigenerated}
\bibfield{author}{\bibinfo{person}{Yiluo Wei} {and} \bibinfo{person}{Gareth Tyson}.} \bibinfo{year}{2024}\natexlab{}.
\newblock \showarticletitle{Understanding the {Impact} of {AI}-{Generated} {Content} on {Social} {Media}: {The} {Pixiv} {Case}}. In \bibinfo{booktitle}{\emph{Proceedings of the 32nd {ACM} {International} {Conference} on {Multimedia}}} \emph{(\bibinfo{series}{{MM} '24})}. \bibinfo{publisher}{Association for Computing Machinery}, \bibinfo{address}{New York, NY, USA}, \bibinfo{pages}{6813--6822}.
\newblock
\showISBNx{979-8-4007-0686-8}
\href{https://doi.org/10.1145/3664647.3680631}{doi:\nolinkurl{10.1145/3664647.3680631}}


\bibitem[Weld et~al\mbox{.}(2022)]%
        {weld_what_2022}
\bibfield{author}{\bibinfo{person}{Galen Weld}, \bibinfo{person}{Amy~X. Zhang}, {and} \bibinfo{person}{Tim Althoff}.} \bibinfo{year}{2022}\natexlab{}.
\newblock \showarticletitle{What {Makes} {Online} {Communities} ‘{Better}’? {Measuring} {Values}, {Consensus}, and {Conflict} across {Thousands} of {Subreddits}}.
\newblock \bibinfo{journal}{\emph{Proceedings of the International AAAI Conference on Web and Social Media}}  \bibinfo{volume}{16} (\bibinfo{date}{May} \bibinfo{year}{2022}), \bibinfo{pages}{1121--1132}.
\newblock
\showISSN{2334-0770}
\href{https://doi.org/10.1609/icwsm.v16i1.19363}{doi:\nolinkurl{10.1609/icwsm.v16i1.19363}}


\bibitem[Wohn(2019)]%
        {Wohn19}
\bibfield{author}{\bibinfo{person}{Donghee~Yvette Wohn}.} \bibinfo{year}{2019}\natexlab{}.
\newblock \showarticletitle{Volunteer Moderators in Twitch Micro Communities: How They Get Involved, the Roles They Play, and the Emotional Labor They Experience}. In \bibinfo{booktitle}{\emph{Proceedings of the 2019 CHI Conference on Human Factors in Computing Systems}} (Glasgow, Scotland Uk) \emph{(\bibinfo{series}{CHI '19})}. \bibinfo{publisher}{Association for Computing Machinery}, \bibinfo{address}{New York, NY, USA}, \bibinfo{pages}{1–13}.
\newblock
\showISBNx{9781450359702}
\href{https://doi.org/10.1145/3290605.3300390}{doi:\nolinkurl{10.1145/3290605.3300390}}


\bibitem[Wojtowicz and DeDeo(2025)]%
        {wojtowiczUnderminingMentalProof2025}
\bibfield{author}{\bibinfo{person}{Zachary Wojtowicz} {and} \bibinfo{person}{Simon DeDeo}.} \bibinfo{year}{2025}\natexlab{}.
\newblock \showarticletitle{Undermining {Mental} {Proof}: {How} {AI} {Can} {Make} {Cooperation} {Harder} by {Making} {Thinking} {Easier}}.
\newblock \bibinfo{journal}{\emph{Proceedings of the AAAI Conference on Artificial Intelligence}} \bibinfo{volume}{39}, \bibinfo{number}{2} (\bibinfo{date}{April} \bibinfo{year}{2025}), \bibinfo{pages}{1592--1600}.
\newblock
\showISSN{2374-3468}
\href{https://doi.org/10.1609/aaai.v39i2.32151}{doi:\nolinkurl{10.1609/aaai.v39i2.32151}}


\bibitem[Wood(2024)]%
        {woodDarkEchoesExploitative2024}
\bibfield{author}{\bibinfo{person}{Adrian Wood}.} \bibinfo{year}{2024}\natexlab{}.
\newblock \showarticletitle{Dark echoes: {The} exploitative potential of generative {AI} in online harassment}.
\newblock In \bibinfo{booktitle}{\emph{Psybersecurity}}. \bibinfo{publisher}{CRC Press}.
\newblock
\showISBNx{978-1-032-66485-9}
\newblock
\shownote{Num Pages: 40}.


\bibitem[Xue et~al\mbox{.}(2023)]%
        {xue_can_2023}
\bibfield{author}{\bibinfo{person}{Junzhi Xue}, \bibinfo{person}{Lizheng Wang}, \bibinfo{person}{Jinyang Zheng}, \bibinfo{person}{Yongjun Li}, {and} \bibinfo{person}{Yong Tan}.} \bibinfo{year}{2023}\natexlab{}.
\newblock \bibinfo{title}{Can {ChatGPT} {Kill} {User}-{Generated} {Q}\&{A} {Platforms}?}
\newblock
\href{https://doi.org/10.2139/ssrn.4448938}{doi:\nolinkurl{10.2139/ssrn.4448938}}


\bibitem[Yang and Menczer(2024)]%
        {yang23}
\bibfield{author}{\bibinfo{person}{Kai-Cheng Yang} {and} \bibinfo{person}{Filippo Menczer}.} \bibinfo{year}{2024}\natexlab{}.
\newblock \showarticletitle{Anatomy of an {AI}-powered malicious social botnet}.
\newblock \bibinfo{journal}{\emph{Journal of Quantitative Description: Digital Media}}  \bibinfo{volume}{4} (\bibinfo{year}{2024}).
\newblock
\href{https://doi.org/10.51685/jqd.2024.icwsm.7}{doi:\nolinkurl{10.51685/jqd.2024.icwsm.7}}
\newblock
\shownote{Section: ICWSM 2024 Special Issue}.


\bibitem[Yang et~al\mbox{.}(2024)]%
        {yang_survey_2023}
\bibfield{author}{\bibinfo{person}{Xianjun Yang}, \bibinfo{person}{Liangming Pan}, \bibinfo{person}{Xuandong Zhao}, \bibinfo{person}{Haifeng Chen}, \bibinfo{person}{Linda~Ruth Petzold}, \bibinfo{person}{William~Yang Wang}, {and} \bibinfo{person}{Wei Cheng}.} \bibinfo{year}{2024}\natexlab{}.
\newblock \showarticletitle{A {Survey} on {Detection} of {LLMs}-{Generated} {Content}}. In \bibinfo{booktitle}{\emph{Findings of the {Association} for {Computational} {Linguistics}: {EMNLP} 2024}}. \bibinfo{address}{Miami, Florida, USA}, \bibinfo{pages}{9786--9805}.
\newblock
\href{https://doi.org/10.18653/v1/2024.findings-emnlp.572}{doi:\nolinkurl{10.18653/v1/2024.findings-emnlp.572}}


\bibitem[Zhang et~al\mbox{.}(2020)]%
        {Zhang20}
\bibfield{author}{\bibinfo{person}{Amy~X. Zhang}, \bibinfo{person}{Grant Hugh}, {and} \bibinfo{person}{Michael~S. Bernstein}.} \bibinfo{year}{2020}\natexlab{}.
\newblock \showarticletitle{PolicyKit: Building Governance in Online Communities}. In \bibinfo{booktitle}{\emph{Proceedings of the 33rd Annual ACM Symposium on User Interface Software and Technology}} (Virtual Event, USA) \emph{(\bibinfo{series}{UIST '20})}. \bibinfo{publisher}{Association for Computing Machinery}, \bibinfo{address}{New York, NY, USA}, \bibinfo{pages}{365–378}.
\newblock
\showISBNx{9781450375146}
\href{https://doi.org/10.1145/3379337.3415858}{doi:\nolinkurl{10.1145/3379337.3415858}}


\bibitem[Zhang et~al\mbox{.}(2022)]%
        {Zhang22}
\bibfield{author}{\bibinfo{person}{Lei Zhang}, \bibinfo{person}{Tianying Chen}, \bibinfo{person}{Olivia Seow}, \bibinfo{person}{Tim Chong}, \bibinfo{person}{Sven Kratz}, \bibinfo{person}{Yu~Jiang Tham}, \bibinfo{person}{Andr\'{e}s Monroy-Hern\'{a}ndez}, \bibinfo{person}{Rajan Vaish}, {and} \bibinfo{person}{Fannie Liu}.} \bibinfo{year}{2022}\natexlab{}.
\newblock \showarticletitle{Auggie: Encouraging Effortful Communication through Handcrafted Digital Experiences}.
\newblock \bibinfo{journal}{\emph{Proc. ACM Hum.-Comput. Interact.}} \bibinfo{volume}{6}, \bibinfo{number}{CSCW2}, Article \bibinfo{articleno}{427} (\bibinfo{date}{Nov.} \bibinfo{year}{2022}), \bibinfo{numpages}{25}~pages.
\newblock
\href{https://doi.org/10.1145/3555152}{doi:\nolinkurl{10.1145/3555152}}


\end{thebibliography}

\end{document}